\documentclass{emulateapj}
\usepackage{lscape}

\newcommand{\ee}[1]{\mbox{${} \times 10^{#1}$}}% scientific number format
\newcommand{\degree}{\mbox{$^{\circ}$}}

\newcommand{\as}{\mbox{\arcsec}}
\newcommand{\um}{$\mu$m}
\newcommand{\lsun}{\mbox{L$_\odot$}}% Lsun
\newcommand{\msun}{\mbox{M$_\odot$}}% Msun

\newcommand{\lbol}{\mbox{$L_{bol}$}} % bolometric luminosity
\newcommand{\tbol}{\mbox{$T_{bol}$}} % bolometric temperature
\newcommand{\lbolsmm}{\mbox{$L_{bol}/L_{smm}$}} % bol-to-smm luminosity

\newcommand{\lsmm}{\mbox{$L_{smm}$}} % luminosity longward of 350 mic.
\input{epsf}

\newcommand{\andre}{Andr\'{e}}

\slugcomment{\footnotesize Accepted for Publication in ApJ}
\begin{document}
%%%%%%%%%%%%%%%%%% title %%%%%%%%%%%%%%%%%%%%%%%%%%%%%%%%%%%%%%%%
\title {\bf Evolutionary Signatures in the Formation of Low-Mass Protostars. II. Towards Reconciling Models and Observations}
\author{
Michael M.~Dunham\altaffilmark{1,2}, Neal J.~Evans II\altaffilmark{1}, Susan Terebey\altaffilmark{3}, Cornelis P.~Dullemond\altaffilmark{4}, \& Chadwick H.~Young\altaffilmark{5}
}
\altaffiltext{1}{Department of Astronomy, The University of Texas at Austin, 1 University Station, C1400, Austin, Texas 78712--0259, USA}

\altaffiltext{2}{mdunham@astro.as.utexas.edu}

\altaffiltext{3}{Department of Physics and Astronomy PS315, 5151 State University Drive, California State University at Los Angeles, Los Angeles, CA 90032, USA}

\altaffiltext{4}{Max Planck Institut f\"{u}r Astronomie, K\"{o}nigstuhl 17, 69117 Heidelberg, Germany}

\altaffiltext{5}{Department of Physical Sciences, Nicholls State University, Thibodaux, LA 70310, USA}

\begin{abstract}

A long-standing problem in low-mass star formation is the ``luminosity problem,'' whereby protostars are underluminous compared to the accretion luminosity expected both from theoretical collapse calculations and arguments based on the minimum accretion rate necessary to form a star within the embedded phase duration.  Motivated by this luminosity problem, we present a set of evolutionary models describing the collapse of low-mass, dense cores into protostars.  We use as our starting point the evolutionary model following the inside-out collapse of a singular isothermal sphere as presented by Young \& Evans (2005).  We calculate the radiative transfer of the collapsing core throughout the full duration of the collapse in two dimensions.  From the resulting spectral energy distributions, we calculate standard observational signatures (\lbol, \tbol, \lbolsmm) to directly compare to observations.  We incorporate several modifications and additions to the original Young \& Evans model in an effort to better match observations with model predictions: (1) we include the opacity from scattering in the radiative transfer, (2) we include a circumstellar disk directly in the two-dimensional radiative transfer, (3) we include a two-dimensional envelope structure, taking into account the effects of rotation, (4) we include mass-loss and the opening of outflow cavities, and (5) we include a simple treatment of episodic mass accretion.  We find that scattering, two-dimensional geometry, mass-loss, and outflow cavities all affect the model predictions, as expected, but none resolve the luminosity problem.  On the other hand, we find that a cycle of episodic mass accretion similar to that predicted by recent theoretical work can resolve this problem and bring the model predictions into better agreement with observations.  Standard assumptions about the interplay between mass accretion and mass loss in our model give star formation efficiencies consistent with recent observations that compare the core mass function (CMF) and stellar initial mass function (IMF).  Finally, the combination of outflow cavities and episodic mass accretion reduce the connection between observational Class and physical Stage to the point where neither of the two commonly used observational signatures (\tbol\ and \lbolsmm) can be considered reliable indicators of physical Stage.
\end{abstract}

%%%%%%%%%%%%%%%%%%%%%%%%%%%%%%%%%%%%%%%%%%%%%%%%%%%%%%%%%%%%%

\section{Introduction}\label{intro}

Over the past few decades a general picture of low-mass star formation has emerged.  As first presented by Adams et al.~(1987) and summarized by Shu, Adams, \& Lizano (1987), this picture merges an empirical classification scheme based on the infrared spectral slope (Lada \& Wilking 1984) with a theory involving the stages of the collapse of a dense, rotating core (Shu 1977; Terebey, Shu, \& Cassen 1984, hereafter TSC84).  In Stage I, the core begins collapsing and the newly formed protostar\footnote{We adopt the definition of a protostar as the central object within a core collapsing to form a star.} is initially heavily obscured by the surrounding envelope, exhibiting a Class I spectral energy distribution (SED) rising from 2 to 20 \um\ due to reprocessing of short-wavelength emission by the dust in the envelope.  Conservation of angular momentum causes a disk to build up (e.g., Adams \& Shu 1986).  The envelope dissipates through accretion and mass-loss processes.  Once it fully dissipates the object transitions from Stage I to Stage II, leaving a pre-main sequence star surrounded by a circumstellar disk that exhibits a Class II SED falling from 2 to 20 \um, but with a shallower slope than expected for a main-sequence star due to ``extra'' infrared emission from the dust in the disk.  The disk eventually dissipates, leaving a Stage III pre-main sequence star exhibiting a Class III SED consistent (or at least nearly so; see Evans et al.~[2009] and references therein) with that expected for a main-sequence star.  \andre\ et al.~(1993) later added Class 0 to the scheme, defining such objects observationally as emitting a relatively large fraction (greater than $0.5$\%) of their total luminosity at wavelengths $\lambda \geq 350$ \um.  Defining a corresponding physical stage, Stage 0 objects are young, embedded protostars with greater than 50\% of their total system mass still in the envelope (\andre\ et al.~1993).  While the term ``Class'' is often assumed to have both meanings in the literature, in this work we follow Robitaille et al.~(2006) and distinguish between ``Class'', determined by observed quantities, and ``Stage'', determined by the ratio of envelope mass to total system mass.

Despite the successes of this picture many questions remain, including a detailed understanding of the mass accretion process from the core to the star.  The ``standard model'' of star formation, the inside-out collapse of an isothermal sphere calculated by Shu (1977) and extended by TSC84 to include rotation, predicts a constant mass accretion rate of about 2\ee{-6} \msun\ yr$^{-1}$.  This gives rise to the classic ``luminosity problem'' whereby accretion at such a rate produces accretion luminosities ($L_{acc} \propto M_* \dot{M}$) higher than typically observed for embedded protostars.  First noticed by Kenyon et al.~(1990), the problem has recently been emphasized by results from the \emph{Spitzer Space Telescope}.  Dunham et al.~(2008), Enoch et al.~(2009), and Evans et al.~(2009) all show that the distribution of embedded protostellar luminosities is strongly peaked at low luminosities.  Enoch et al.~and Evans et al.~both find that a substantial fraction (greater than $50$\%) of embedded protstars have luminosities suggesting accretion rates $\dot{M} \la 10^{-6}$ \msun\ yr$^{-1}$, and Dunham et al.~argue that the large fraction of sources at low luminosities is inconsistent with a constant mass accretion rate.

To compare theoretical models of star formation to observations, Young \& Evans (2005; hereafter YE05) used a one-dimensional dust radiative transfer package to calculate the observational signatures of cores undergoing inside-out collapse following Shu (1977).  They followed three different cores with initial masses of 0.3, 1, and 3 \msun\ from the onset of collapse until the end of the embedded phase, calculating the bolometric luminosity (\lbol), bolometric temperature (\tbol), and ratio of bolometric to submillimeter luminosity (\lbolsmm, see \S \ref{evolsigs}).  \tbol\ is defined by Myers \& Ladd (1993) as the temperature of a blackbody with the same flux-weighted mean frequency as the source (see \S \ref{evolsigs}), and can be thought of as a protostellar equivalent of $T_{eff}$; \tbol\ starts at very low values ($\sim 10$ K) for cold, starless cores and eventually increases to $T_{eff}$ once the envelope and disk have fully dissipated.  YE05 compared their model to observations by plotting both their model tracks and observations of sources on a plot of \lbol\ vs. \tbol, which Myers et al.~(1998) called a BLT diagram.  This figure (Figure 19 in YE05) shows that observed sources at a given \tbol\ range from having \lbol\ consistent with the Young \& Evans model tracks to having \lbol\ up to $1-3$ orders of magnitude lower than predicted, clearly illustrating the luminosity problem.

An idea proposed to resolve the luminosity problem is that mass accretion is episodic in nature, and the protostars with the lowest luminosities are those observed in quiescent accretion states (e.g., Kenyon et al.~1990, Kenyon \& Hartmann 1995; YE05; Enoch et al.~2009; Evans et al.~2009).  Theoretical studies have provided several mechanisms by which such a process may occur, such as material piling up in a circumstellar disk until gravitational instabilities drive angular momentum outward and mass inward in short-lived bursts (Vorobyov \& Basu 2005b, 2006; Boss 2002).  Alternatively, accretion bursts may be driven by a combination of gravitational and magnetorotational instabilities (Zhu et al.~2009), or quasi-periodic magnetically driven outflows in the envelope may cause mass accretion onto the protostar to occur in magnetically controlled bursts (Tassis \& Mouschovias 2005).  Indeed, observational evidence for non-steady mass accretion in young protostellar systems still in the embedded phase now exists in the form of accretion bursts in Class I sources (e.g., Acosta-Pulido et al.~2007; K\'{o}sp\'{a}l et al.~2007; Fedele et al.~2007) and Class 0 sources with strong outflows implying higher average mass accretion rates than expected from currently observed low luminosities (e.g., Dunham et al.~2006; \andre\ et al.~1999; M.~M.~Dunham et al.~2009, in preparation).  Additionally, Watson et al. (2007) showed a mismatch between the accretion rates onto the disk and protostar of NGC 1333-IRAS 4B (measured by modeling water emission lines and by assuming all of the observed luminosity is accretion luminosity, respectively), a result they have now expanded to other sources (D.M. Watson et al.~[2009], in preparation).  Finally, episodic mass ejection is seen in jets ejected from some protostellar systems, suggesting an underlying variability in the mass accretion, although the combination of jet velocities and spacing between knots often suggests shorter periods of episodicity than found by the above theoretical studies.  For example, Lee et al.~(2007) found a period of $\sim 15-44$ yr for the periodic protostellar jet HH 211.  We also note here that an alternative collapse scenario, ``outside-in'' collapse, where the collapse is triggered by an external shock wave, can produce a range of mass accretion rates roughly in agreement with those derived from observations and predicted by episodic accretion models (Boss 1995).  However, while such a process is relevant for massive star-forming regions and possibly for the formation of our own solar system (Boss 2008), it is not likely relevant for the relatively isolated protostars forming in most nearly, low-mass star forming regions.

Here we will test the hypothesis that episodic accretion can solve the luminosity problem.  First, however, we will test the effects of several other possibilities that were not included in the YE05 model, including revised dust opacities, two-dimensional disk and envelope geometry, and mass-loss and outflow cavities.  This work is complementary to several other recent modeling efforts.  Myers et al.~(1998) included the effects of mass-loss in their calculations of the evolution of \lbol\ and \tbol\ with time for collapsing cores, but they did not include outflow cavities and their model evolution is not based on a fully self-consistent model such as the collapse solutions calculated by Shu (1977) or TSC84.  Whitney et al.~(2003a, 2003b), Robitaille et al.~(2006), and Crapsi et al.~(2008) all used 2-D radiative transfer models to investigate the effects of two-dimensional disk and envelope geometry and outflow cavities on the evolutionary signatures of embedded protostars.  However, none of these authors present a full evolutionary model following the collapse of a core but instead vary parameters over pre-defined grids to capture typical protostars of different evolutionary stages, and only Crapsi et al.~(2008) considered the predictions of observed quantities other than infrared colors.  Lee (2007) included episodic accretion in the YE05 model in a very simple manner in order to study the effects such a process has on the chemical evolution of collapsing cores.  Myers (2008) presented an analytic calculation of the growth of a protostar through competing infall and dispersal processes; some aspects of our models, in particular the opening of outflow cavities, are similar to those featured by Myers (2008).  Baraffe, Chabrier, \& Gallardo (2009) showed that episodic accretion in the early, embedded phase can explain the observed luminosity spread in H$-$R diagrams of star-forming regions at a few Myr without having to invoke large age spreads.  Finally, Vorobyov (2009b) compared the distribution of mass accretion rates in their simulations of episodic accretion (Vorobyov \& Basu 2005b, 2006) to those inferred from the luminosities of protostars in the Perseus, Serpens, and Ophiuchus molecular clouds compiled by Enoch et al.~(2009) and showed that their simulations reproduced some of the basic features of the observed distribution of mass accretion rates.

In this paper we revisit the YE05 model, which is summarized in \S \ref{model}.  Following YE05, we assume a distance of 140 pc for all models to calculate observed SEDs.  This assumed distance only affects the absolute flux level when we display SEDs, all other observational signatures are independent of the assumed distance.  We discuss the method we use to compare evolutionary models to observations in \S \ref{observations}.  In \S \ref{modifications}, we make several updates and additions to the model in a step-by-step fashion, examining the results of each modification individually.  Specifically, we include scattering in the radiative transfer calculations in \S \ref{mod1}.  In \S \ref{mod2}$-$\ref{mod3} we generalize the model from its original, one-dimensional structure to two dimensions with a more realistic disk (\S \ref{mod2}) and two-dimensional envelope structure (\S \ref{mod3}).  We include the effects of mass-loss and outflow cavities in \S \ref{mod4}, and in \S \ref{mod5} we include a simple treatment of episodic accretion.  A discussion of the results of this work is presented in \S \ref{discussion}, and we present a summary of our conclusions in \S \ref{summary}.  We note here that choices of parameters in the models presented below are physically motivated and theoretically and/or observationally constrained whenever possible.  However, these are simple, idealized models that are not always fully self-consistent.  Limitations are discussed as each modification is described.

\section{Description of the Original Model}\label{model}

We begin with a summary of the YE05 model.  We provide a fairly comprehensive description of this model to place our work in later sections in context, but refer the reader to YE05 for a complete description.

\subsection{Evolution of the Envelope, Protostar, and Disk}\label{model_setup}

The YE05 model follows the collapse of singular isothermal spheres with initial masses of 0.3, 1, and 3 \msun\ according to the inside-out collapse solution calculated by Shu (1977).  This model begins with an envelope radial density profile proportional to $r_{env}^{-2}$, truncated at an outer radius that sets the initial core mass (YE05 Equation 1)\footnote{Following the convention adopted by YE05, radii pertaining to the envelope are denoted by lowercase r, while radii pertaining to either the star or disk are denoted by uppercase R.}.  The collapse of the envelope begins at the center and moves outward at the sound speed, giving rise to an infall radius that moves outward with time.  The density profile is then described approximately as a broken power law; inside the infall radius $n \propto r_{env}^{-3/2}$, indicative of free-fall, while outside the infall radius the profile remains the initial $n \propto r_{env}^{-2}$.  YE05 used the exact solutions from Shu (1977) to account for the transition region between the two.  Once the infall radius exceeds the envelope outer radius, the model adopts a density profile with $n \propto r_{env}^{-3/2}$ everywhere.  The envelope inner radius is held fixed at a value such that the initial optical depth at 100 \um\ is set equal to 10 (YE05 Equation 4; see YE05 for a discussion of the effects of varying this initial optical depth) until the disk outer radius exceeds this value (see below); once this occurs the inner envelope radius is set equal to the disk outer radius.

The mass of the envelope, $M_{env}$, declines as mass accretes from the envelope to the protostar+disk system at the rate $\dot{M}_{env}$.  In the Shu (1977) solution, this rate is constant and given by 
\begin{equation}\label{eq_mdot1}
\dot{M}_{env} = m_0 \frac{c_s^3}{G} \qquad ,
\end{equation}
where $m_0$ is a dimensionless constant of order unity, $G$ is the gravitational constant, and $c_s$ is the effective sound speed including thermal and turbulent components and calculated by YE05 as
\begin{equation}\label{eq_aeff1}
c_s = \left( \frac{kT}{\mu m_H} + \frac{1}{2} v_{turb}^2 \right)^{1/2} \qquad .
\end{equation}
With $k$ the Boltzmann constant, $T$ the isothermal temperature (assumed to be 10 K), $\mu$ the mean molecular mass (2.29 for a gas that is 25\% by mass helium), $m_H$ the mass of the hydrogen atom, and $v_{turb}$ the turbulent velocity, chosen such that the thermal and turbulent contributions are equal, $c_s = 0.268$ km s$^{-1}$.  This gives an envelope mass accretion rate of 4.57 $\times$ 10$^{-6}$ \msun\ yr$^{-1}$. This is 5\% lower than the YE05 value of $\dot{M}_{env} =$ 4.8 $\times$ 10$^{-6}$ \msun\ yr$^{-1}$; this difference arises from correcting a small numerical error in the YE05 model.

To include a protostellar disk in their 1-D model, YE05 use the method developed by Butner et al.~(1994), based on the disk model of Adams et al.~(1988).  This method simulates a disk by calculating the emission from a disk with given surface density and temperature profiles at a given inclination, averaging over all inclinations, and then adding this average spectrum to the (proto)stellar spectrum to form the final input spectrum of the internal source for the 1-D radiative transfer calculation through the envelope.  Both the surface density and temperature profiles are described as power laws: $\Sigma(R_{disk}) \propto R_{disk}^{-p}$ and $T(R_{disk}) \propto R_{disk}^{-q}$.  YE05 choose $p$ = 1.5, following Butner et al.~(1994) and Chiang \& Goldreich (1997), and $q$ = 0.5 to simulate a flared disk.

The inner radius of the disk is set equal to the dust destruction radius, calculated (assuming spherical, blackbody dust grains) as
\begin{equation}
R_{disk}^{in} = \sqrt{\frac{L_*}{16 \pi \sigma T_{dust}^4}} \qquad ,
\end{equation}
where $L_*$ is the protostellar luminosity (see below) and $T_{dust}$ is the dust destruction temperature.  YE05 assume $T_{dust}$ = 2000 K; here we update this value to $T_{dust}$ = 1500 K (e.g., Cieza et al.~2005).  The outer radius of the disk is set to the centrifugal radius, which increases with time as (TSC84)
\begin{equation}
R_{disk}^{out} = \frac{m_0^3}{16} c_s t^3 \Omega_0^2 \qquad ,
\end{equation}
where $t$ is the time since the onset of the collapse and $\Omega_0$ is the initial angular velocity of the cloud.  YE05 set $\Omega_0$ such that the disk outer radius is 100 AU at the end of the collapse of each core ($\Omega_0$ = 3.4\ee{-13}, 5.5\ee{-14}, and 1.0\ee{-14} s$^{-1}$ for the 0.3, 1, and 3 \msun\ cores, respectively).  The 5\% lower envelope accretion rate from YE05 results in 5\% longer total collapse duration, but since we choose to keep the values of $\Omega_0$ set by YE05 to minimize changes and facilitate direct comparison between their results and the results of this work, the final disk outer radii are approximately 15\% larger than 100 AU.

Following Adams \& Shu (1986), YE05 assume that all mass accreted from the envelope accretes onto either the star or the disk at rates $\dot{M}_*$ and $\dot{M}_{disk}$, where $\dot{M}_{env} = \dot{M}_* + \dot{M}_{disk}$.  These rates are governed by $u_*$, the ratio of the protostellar and disk outer radii ($u_* = R_*/R_{disk}^{out}$).  With the protostellar radius calculated following Palla \& Stahler (1991), except at early times ($t < 2 \times 10^4$ yr) where it is set to 5 AU to simulate the First Hydrostatic Core (Masunaga et al.~1998; Boss \& Yorke 1995; see YE05 for details), Adams \& Shu use the velocity field and density profile of the collapse solution for a rotating, singular isothermal sphere (Cassen \& Moosman 1981; TSC84) to determine the flux of material flowing from the cloud directly onto the protostar and disk and calculate the protostellar and disk mass accretion rates as

\begin{equation}
\dot{M}_* = \dot{M}_{env} [ 1 - (1 - u_*)^{1/2}] \qquad ,
\end{equation}
\begin{equation}
\dot{M}_{disk} = \dot{M}_{env} ( 1 - u_*)^{1/2} \qquad .
\end{equation}

In addition to direct accretion from the envelope (which quickly becomes negligible as the disk outer radius grows and thus $u_*$ decreases), the protostar also accretes mass from the disk at the rate $\dot{M}_{DtoP} = \eta_D \dot{M}_{env}$, where $\eta_D$ is an efficiency factor assumed to be 0.75 (see YE05 for a discussion of the effects of different assumed values for $\eta_D$).  Following Adams \& Shu (1986), YE05 calculate the mass of the disk, $M_{disk}$, including both accretion from the envelope and onto the protostar (see YE05 Equations 12 and 13).  The mass of the protostar is then calculated as $M_* = M_{int} - M_{disk}$, where $M_{int}$ is the total internal mass accreted from the envelope ($M_{int} = \dot{M}_{env}t$).

\subsection{Luminosity Sources}\label{model_luminosities}

The total internal luminosity of the protostar and disk at each point in the collapse from core to star contains several components.  Following Adams \& Shu (1986), YE05 include six components:

\begin{enumerate}
\item $L_{EtoP}$: luminosity arising from accretion from the envelope directly onto the protostar.
\item $L_{EtoD}$: luminosity arising from accretion from the envelope onto the disk.
\item $L_{DtoP}$: luminosity arising from accretion from the disk onto the protostar.
\item $L_{DM}$: disk ``mixing luminosity'' arising from luminosity released when newly accreted material with its own radial and angular velocity components mixes with disk material in a Keplerian orbit, putting the new and old material into a new Keplerian orbit (see Adams \& Shu [1986] for details).
\item $L_{DR}$: luminosity arising from the release of energy stored in differential rotation of the protostar.
\item $L_{phot}$: luminosity arising from gravitational contraction and deuterium burning.
\end{enumerate}

The first two components are both directly proportional to $\dot{M}_{env}$, with geometrical factors that depend on $u_*$ to account for the changing rates of accretion onto the protostar and disk (see Equations 27 and 29a of Adams \& Shu 1986 for the exact definitions of each term).  Both components are quite small throughout the collapse of each core; $L_{EtoP}$ since the amount of material accreting directly from the envelope to the star becomes small very quickly, and $L_{EtoD}$ because $R_{disk}^{out} >> R_*$ (except for very early times) and thus the material accreting onto the disk has not yet fallen very deep into the potential well.

The third component depends on the rate at which material accretes from the disk to the protostar, controlled by the efficiency factor $\eta_D$.  The exact definition is given by Equation 30 of Adams \& Shu (1986); it is essentially just one-half of the spherical accretion luminosity arising from material accreting at this rate (the other half of the initial gravitational potential energy is stored as kinetic energy of the disk material's Keplerian rotation shortly before it accretes onto the star), with both the luminosity already released from accretion onto the disk and from the disk mixing (see below) subtracted out.  This is the dominant source of luminosity once the disk has formed.  Any luminosity arising from the inward transport of material within the disk is indirectly included in this term since its calculation starts with the total spherical accretion luminosity.

The fourth component is also directly proportional to $\dot{M}_{env}$, with geometrical factors that depend on $u_*$.  The exact definition is given by Equation 29b of Adams \& Shu (1986).

The fifth component depends on the total rate of accretion onto the star and the efficiency $\eta_*$ with which energy stored in differential rotation of the protostar is released (assumed to be $\eta_* = 0.5$; see YE05 for details).  The exact definition is given by Equation 32 of Adams \& Shu (1986).

To include the sixth component, the luminosity arising from gravitational contraction and deuterium burning, YE05 used the pre-main sequence tracks of D'Antona \& Mazzitelli (1994) with opacities from Alexander et al.~(1989).  They also assumed a power-law expression to extrapolate to times earlier than those included in the pre-main sequence tracks, $L_{phot} = L_0^{phot} (t/t_0)^5$, where $t_0$ is the earliest time in the tracks and $L_0^{phot}$ is the pre-main sequence luminosity at this time.  Finally, they followed Myers et al.~(1998) in adding $10^5$ yr to the times of the pre-main sequence tracks to account for the delay between the onset of collapse and the ``zero time'' of these tracks.  Thus it is only at late times in the collapse of the cores that $L_{phot}$ becomes an important source of luminosity.

Finally, there is also external luminosity arising from heating of the envelope by the Interstellar Radiation Field (ISRF).  We adopt the same ISRF as YE05, and input the mean intensity of the ISRF ($J_{ext}$) into the radiative transfer code as an additional source of heating.  The luminosity added to \lbol\ from this external heating, $L_{ext}$, is determined after each radiative transfer model is run by subtracting the total internal luminosity (the sum of the above six components) from the total model luminosity.

\section{Comparing Models to Observations}\label{observations}

For all of the models presented below, we use the two-dimensional, axisymmetric, Monte Carlo dust radiative transfer package RADMC (Dullemond \& Turolla 2000; Dullemond \& Dominik 2004) to calculate the two-dimensional dust temperature profile of the envelope at each timestep throughout the collapse of the 0.3, 1, and 3 \msun\ cores (following YE05, $\Delta t =$ 1000, 2000, and 6000 yr for the 0.3, 1, and 3 \msun\ cores, respectively).  Spectral Energy Distributions (SEDs) at each timestep are then calculated at 9 different inclinations ranging from $i=5-85$\degree\ in steps of 10\degree\ except for model 1 (\S \ref{mod1}), where there is no inclination dependence and thus only one SED is calculated per timestep.  An inclination of $i=0$\degree\ corresponds to a pole-on (face-on) system, while an inclination of $i=90$\degree\ corresponds to an edge-on system.

\subsection{Calculating Observational Signatures}\label{evolsigs}

We use the model SEDs to calculate observational signatures of the models at each timestep for each inclination.  We calculate the bolometric luminosity (\lbol), the ratio of bolometric to submillimeter luminosity (\lbolsmm), and the bolometric temperature (\tbol).  \lbol\ is calculated by intergrating over the full SED,
\begin{equation}\label{eq_lbol}
\lbol = 4\pi d^2 \int_0^{\infty} S_{\nu}d\nu \qquad ,
\end{equation}
while the submillimeter luminosity is calculated by integrating over the SED for $\lambda$ $\geq$ 350 \um,
\begin{equation}\label{eq_lsmm}
\lsmm = 4\pi d^2 \int_0^{\nu=c/350\, \mu m} S_{\nu}d\nu \qquad .
\end{equation}
The bolometric temperature is defined to be the temperature of a blackbody with the same flux-weighted mean frequency as the source (Myers \& Ladd 1993).  Following Myers \& Ladd, \tbol\ is calculated as
\begin{equation}\label{eq_tbol}
\tbol = 1.25 \times 10^{-11} \, \frac{\int_0^{\infty} \nu S_{\nu} d\nu}{\int_0^{\infty} S_{\nu} d\nu} \quad \rm{K} \qquad .
\end{equation}
The integrals defined in equations \ref{eq_lbol} $-$ \ref{eq_tbol} are calculated using the trapezoid rule to integrate the finitely sampled model SEDs.

Both \tbol\ and \lbolsmm\ can be used as alternatives to the infrared spectral slope to classify sources.  Evans et al.~(2009) present a comprehensive discussion of source classification by observational signatures, here we briefly summarize the main points.  Chen et al.~(1995) defined class boundaries in \tbol, as follows:

\begin{itemize}
\item[] \textbf{Class 0} \tbol\ $< 70$ K
\item[] \textbf{Class I} $70 \leq$ \tbol\ $\leq 650$ K
\item[] \textbf{Class II} $650 <$ \tbol\ $\leq 2800$ K
\end{itemize}

From the original observational definition of Class 0 by \andre\ et al.~(1993), the class boundaries in \lbolsmm\ are:

\begin{itemize}
\item[] \textbf{Class 0} \lbolsmm\ $\leq 200$
\item[] \textbf{Class I} \lbolsmm\ $> 200$
\end{itemize}

Based on their evolutionary model, YE05 revised the Class 0/I boundary in \lbolsmm\ from 200 to 175.  There is no defined boundary between Class I and Class II in \lbolsmm.

\subsection{Observational Dataset}\label{obsdataset}

We use the 1024 Young Stellar Objects (YSOs) in the five large, nearby molecular clouds surveyed by the \emph{Spitzer Space Telescope} Legacy Project ``From Molecular Cores to Planet Forming Disks'' (Evans et al.~2003) as our observational dataset.  Evans et al.~(2009) compiled photometry and calculated \lbol\ and \tbol\ for all 1024 YSOs in the same manner as described above.  They used both the observed photometry to calculated observed \lbol\ and \tbol\ values, and photometry corrected for foreground extinction to calculate extinction-corrected values of \lbol\ and \tbol\ (see Evans et al.~[2009], and \S \ref{discuss_evolutionary}, for details on the corrections for foreground extinction).  Since evolutionary models have no foreground extinction, only local extinction from the dust in the model itself, we use the extinction-corrected \lbol\ and \tbol.

Evans et al.~(2009) concluded that 112 of the 1024 YSOs are embedded protostars based on association with a millimeter continuum emission source tracing an envelope.  Since the evolutionary models presented both by YE05 and in this paper follow the evolution of protostars up until the point of complete envelope dissipation, we consider these 112 embedded protostars to be our final observational dataset to use when comparing the models to observations.

\subsection{Comparing Models to Observations}\label{comparingmodobs}

With the model observational signatures calculated as described above and the observational dataset of 112 embedded protostars from Evans et al.~(2009), we can compare the model predictions to observations.

\begin{figure}[hbt!]
\epsscale{0.90}
\plotone{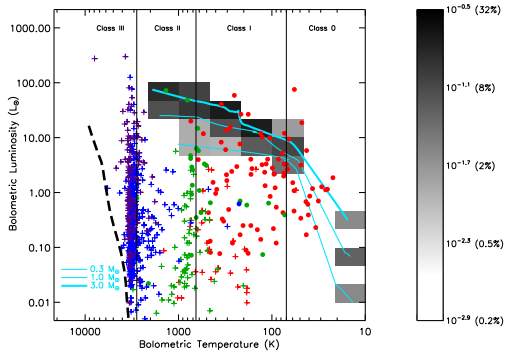}
\caption{\label{fig_modye05_blt}BLT diagram for the YE05 model.  The light blue lines show \lbol\ $-$ \tbol\ tracks for the 0.3, 1, and 3 \msun\ cores.  The grayscale pixels indicate the fraction of total time the model spends in each \lbol\ $-$ \tbol\ bin, calculated from Equation \ref{eq_bins}.  The grayscale is displayed in a logarithmic stretch with the scaling chosen to emphasize the full extent of the models in \lbol\ $-$ \tbol\ space.  The mapping between grayscale and fraction of total time is indicated in the legend.  The class boundaries in \tbol\ are taken from Chen et al.~(1995).  The thick dashed line shows the ZAMS (D'Antona \& Mazzitelli 1994) from 0.1 to 2.0 \msun.  The colored symbols show the Young Stellar Objects from Evans et al.~(2009) in this diagram; the color indicates spectral class (red for Class 0/I, green for flat spectrum, blue for Class II, and purple for Class III), while a circle or cross indicates the source is or is not, respectively, associated with an envelope as traced by millimeter continuum emission.}
\end{figure}

The most common method of comparing model predictions to observations is by plotting the model tracks on a diagram of \lbol\ vs. \tbol.  This was first done by Myers et al.~(1998), who called such a diagram a BLT diagram.  Figure \ref{fig_modye05_blt} shows the YE05 model tracks for the 0.3, 1, and 3 \msun\ cores plotted on a BLT diagram, similar to Figure 19 from YE05.  Also plotted are the 1024 YSOs from Evans et al.~(2009), with color indicating spectral class (red for Class 0/I, green for Flat spectrum, blue for Class II, and purple for Class III; see Evans et al.~for details) and symbol indicating source type (circles for sources associated with envelopes as traced by millimeter continuum emission, plus signs for sources not associated with envelopes).  The relevant comparison is between the model tracks and the sources plotted as circles.

Plotting \lbol\ $-$ \tbol\ tracks on a BLT diagram such as in Figure \ref{fig_modye05_blt} is an adequate means of comparing model results to observations for the YE05 model.  There is no inclination dependence so there is only one track per core mass, and both \lbol\ and \tbol\ increase monotonically.  However, most models below will have an inclination dependence, making it difficult to compare model tracks to observations when each model has different tracks depending on inclination.  Furthermore, the change in \lbol\ and \tbol\ with time will no longer be monotonic once episodic accretion is introduced (\S \ref{mod5}), eliminating the concept of ``model track'' altogether.

With this in mind, we introduce other methods of comparing to observations.  First, we divide the \lbol\ $-$ \tbol\ space into bins of 1/3 dex in both dimensions.  We then calculate the fraction of total time the model spends in each \lbol\ $-$ \tbol\ bin ($f_{bin}$) as follows:

\begin{equation}\label{eq_bins}
f_{bin} = \frac{\displaystyle\sum_{mass} \left( \displaystyle\sum_{inc} t_{inc}w_{inc} \right) w_{mass}}{\displaystyle\sum_{mass} \left( \displaystyle\sum_{inc} t_{collapse}w_{inc} \right) w_{mass}} \qquad ,
\end{equation}
where the numerator is the time spent in the bin and the denominator is the total time.  The interior sum in both the numerator and denominator is over the 9 different inclinations while the exterior sum is over the three different initial core masses.  The quantity $t_{inc}$ is the total time that the SED at a given inclination spends in the specified \lbol\ $-$ \tbol\ bin whereas $t_{collapse}$ is the total collapse time of the core (67,000, 224,000, and 673,000 yr for the 0.3, 1, and 3 \msun\ cores, respectively).  $w_{inc}$ is the weight each inclination receives in the sum, defined as the fraction of solid area subtended by that inclination.  This is calculated in practice by assuming each of the 9 SEDs calculated is valid for inclinations spanning the range ($i-5$\degree) to (i+5\degree).

The final quantity in Equation \ref{eq_bins} is $w_{mass}$, the weight given to each of the three initial mass cores.  This is determined by the core mass function (CMF) of starless cores.  Reading directly from the CMF plot presented by Enoch et al.~(2008; their Figure 13), we find a ratio of 1 to 3 and 1 to 0.3 \msun\ cores of $N_1/N_3 = 2.3$ and $N_1/N_{0.3} = 2.8$.  Alternatively, using their best-fit power-law of $dN/DM \propto M^{-2.3 \pm 0.4}$ gives\footnote{As the power-law is only fit to the CMF for $M>0.8$ \msun, it is inappropriate to use it to obtain an estimate of $N_1/N_{0.3}$} $N_1/N_3 = 3.9^{+2.1}_{-1.4}$, while using their best-fit lognormal distribution gives $N_1/N_3 = 1.1$ and $N_1/N_{0.3} = 12.9$.  If we instead use the three-component power-law IMF found by Kroupa (2002) and assume a 30\% star-formation efficiency in dense gas (Alves et al.~2007; see also \S \ref{mod4}) to scale from the IMF to the CMF, we obtain $N_1/N_3 = 2.3$ and $N_1/N_{0.3} = 0.6$.  Finally, if we assume the same efficiency but instead use the IMF found by Muench et al.~(2002) for the Trapezium cluster shown by Alves et al.~(2007) to match (with the 30\% scaling) the dense core mass function in the pipe nebula, we find $N_1/N_3 = 1.1$ and $N_1/N_{0.3} = 3.5$.  Given the uncertainties in determining the exact CMF, both from uncertainties in core mass calculations and from completeness effects (see Enoch et al.~2008 for a complete discussion), we simply average the above values\footnote{We leave $N_1/N_{0.3} = 12.9$, obtained from the best-fit lognormal distribution in Enoch et al.~(2008), out of the average since it is significantly higher than other values and is derived from a lognormal fit to data that likely suffers from incompleteness effects.} to obtain $N_1/N_3 = 2.1$ and $N_1/N_{0.3} = 2.3$.  Requiring the sum of the weights to be 1 gives $w_{0.3} = 0.2275$, $w_{1} = 0.5233$, and $w_{3} = 0.2492$.

In addition to the model tracks, Figure \ref{fig_modye05_blt} also shows, using grayscale pixels, the fraction of total time the YE05 model spends in each \lbol\ $-$ \tbol\ bin, calculated from Equation \ref{eq_bins} above.  Comparison to the model tracks illustrates that this method of displaying the results has the advantage of showing not only what regions of the BLT diagram the model encompasses but also the relative amount of time spent in different portions of the diagram.  For all of the revised models presented in this paper, we do not show model tracks and instead use only this method of displaying the results on the BLT diagram.

\begin{figure}[hbt!]
\epsscale{1.0}
\plotone{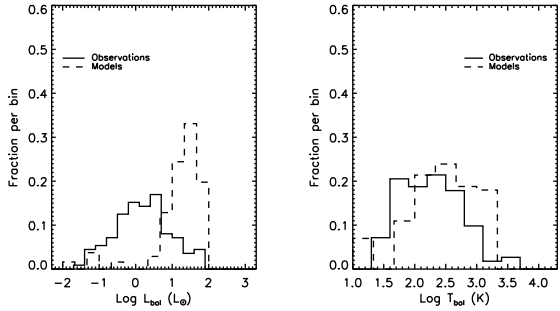}
\caption{\label{fig_modye05_hist}Histograms showing the fraction of total sources (observations; solid lines) and fraction of total time spent by the YE05 model (dashed lines; calculated from Equation \ref{eq_bins}) at various \lbol\ (left) and \tbol\ (right).  The binsize is 1/3 dex in both quantities.  For the observations, only the 112 embedded sources (plotted as filled circles on the BLT diagrams) are included.}
\end{figure}

We can also compare the overall distribution of time the models spend at different values of \lbol\ and \tbol\ to the fraction of total sources observed at those values.  As an example, Figure \ref{fig_modye05_hist} shows \lbol\ and \tbol\ histograms for both the observations and the YE05 model, with binsizes of 1/3 dex for both quantities.  The observational histograms only include the 112 embedded sources associated with envelopes (plotted as filled circles on the BLT diagrams) and plot the fraction of total sources in each bin, while the model histograms plot the fraction of total time spent in each bin calculated from Equation \ref{eq_bins}.  This figure emphasizes the higher luminosities of the model compared to the observations.  We can quantify the agreement between the model and the observations with K-S tests.  Such tests shows that there is less than a 0.1\% probability that the observed and model \lbol\ histograms represent the same underlying distribution, and a 56\% probability that the observed and model \tbol\ histograms represent the same underlying distribution.

\begin{figure}[hbt!]
\epsscale{0.90}
\plotone{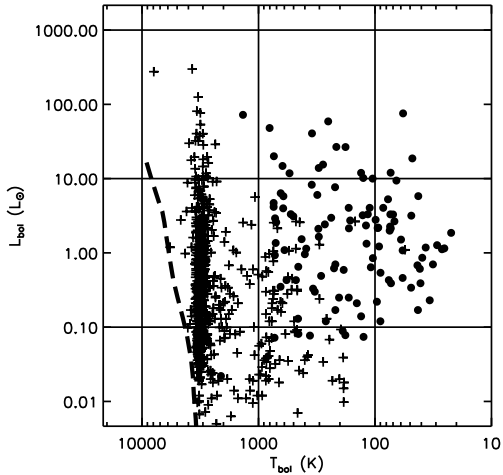}
\caption{\label{fig_modye05_blt2}BLT diagram showing the bins used in the comparison between the fraction of total sources and fraction of total time spent in each bin presented in Table \ref{tab_fractionperbin}.  The thick dashed line shows the ZAMS (D'Antona \& Mazzitelli 1994) from 0.1 to 2.0 \msun.  The symbols show the Young Stellar Objects from Evans et al.~(2009) in this diagram and hold the same meaning as in Figure \ref{fig_modye05_blt}.}
\end{figure}

Finally, we also divide the BLT diagram into much larger bins (1 dex in \tbol\ and 2 dex in \lbol; the bins are shown in Figure \ref{fig_modye05_blt2}).  Column 1 of Table \ref{tab_fractionperbin} lists the \lbol\ and \tbol\ limits of each bin, and column 2 lists the fraction of the 112 embedded sources in each bin.  For comparison, column 3 lists the fraction of total time the YE05 model spends in each bin, calculated from Equation \ref{eq_bins}.  Columns $4-8$ list the same thing for the revised models presented below.  The luminosity problem in the YE05 model is emphasized by the fact that 76.8\% of the observed sources have $0.1 \leq$ \lbol\ $< 10$ \lsun\ while 16.1\% have $10 \leq$ \lbol\ $< 1000$ \lsun, whereas the YE05 model spends only 17.4\% of the time at $0.1 \leq$ \lbol\ $< 10$ \lsun\ but 77.2\% of the time at $10 \leq$ \lbol\ $< 1000$ \lsun.

\section{Modifications to the Original Model}\label{modifications}

We now make several modifications to the YE05 model in a step-by-step fashion, adding them one at a time and discussing the results of each before adding in the next.  To be specific, (1) we include isotropic scattering off dust grains, (2) we include a circumstellar disk directly in the radiative transfer, (3) we include a rotationally flattened  protostellar envelope density structure, (4) we include the effects of mass-loss and outflow cavities, and (5) we include episodic accretion.  These additions are summarized in Table \ref{tab_models}.  All of these additions are made possible by switching to the two-dimensional RADMC rather than the one-dimensional dust radiative transfer package DUSTY (Ivezic et al.~1999) used by YE05.

\subsection{Model 1: Scattering}\label{mod1}

YE05 assumed the dust opacities of Ossenkopf \& Henning (1994) appropriate for thin ice mantles after 10$^5$ yr of coagulation at a gas density of 10$^6$ cm$^{-3}$ (OH5 dust), which several recent studies have shown to be appropriate for cold, dense cores (e.g., Evans et al.~2001; Shirley et al.~2002; Young et al.~2003; Shirley et al.~2005).  These opacities do not extend below 1.25 \um, and they include only the total dust opacity ($\kappa_{tot}$) rather than the contributions from absorption ($\kappa_{abs}$) and scattering ($\kappa_{scat}$) separately.  To remedy this, YE05 also used the dust opacities calculated by Pollack et al.~(1994) for dust grains with a radius of 0.1 \um\ at a temperature of 10 K, which give $\kappa_{abs}$ and $\kappa_{scat}$ separately and extend down to 0.091 \um.  Noting that $\kappa_{tot}$ according to OH5 and Pollack et al.~agreed at short wavelengths, YE05 simply obtained $\kappa_{abs}$ and $\kappa_{scat}$ from Pollack et al.~shortward of 1.25 \um.  Longward of 1.25 \um, they used $\kappa_{tot}$ from the OH5 dust and the albedo (ratio of $\kappa_{scat}/\kappa_{tot}$) from Pollack et al.~to apportion the total OH5 opacity among scattering and absorption components.

Unfortunately, YE05 were not able to include the effects of scattering when using the 1-D radiative transfer code DUSTY to calculate the envelope dust temperature profile and final SED of the protostar+disk+envelope system.  DUSTY assumes that scattering from dust grains is isotropic, when in reality the grains preferentially forward-scatter light.  As described in more detail by YE05, assuming isotropic scattering results in an artifical near-infrared peak in the SED of any core exposed to the ISRF, even a starless core, from backscattering of ISRF photons.  This peak can be as strong as the true peak from thermal dust emission in the far-infrared and submillimeter (YE05).  As the ISRF is the dominant source of heating at early times, YE05 were forced to ignore the effects of scattering in order to produce reasonable results.

\begin{figure}[hbt!]
\epsscale{1.0}
\plotone{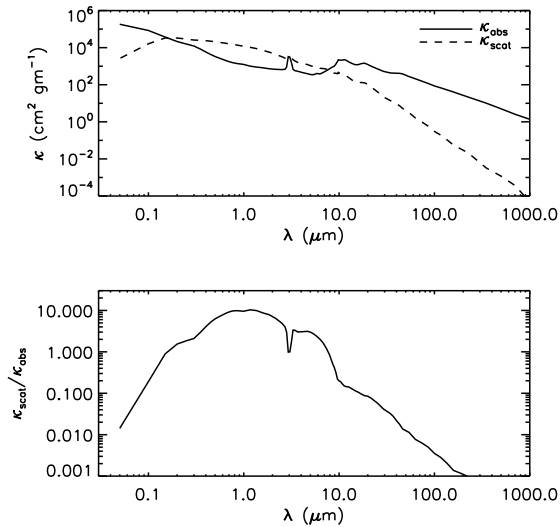}
\caption{\label{fig_scattering}\emph{Top}:  Absorption and scattering opacities ($\kappa_{abs}$ and $\kappa_{scat}$, respectively) constructed by YE05 based on the dust opacity models of Ossenkopf \& Henning (1994) and Pollack et al.~(1994).  \emph{Bottom}:  Ratio of scattering to absorption opacity.}
\end{figure}

However, by ignoring scattering, they removed an important opacity source from their model.  Figure \ref{fig_scattering}, which plots $\kappa_{abs}$, $\kappa_{scat}$, and the ratio of $\kappa_{scat}$ to $\kappa_{abs}$ as a function of wavelength $\lambda$, shows that the opacity from scattering dominates that from absorption over the approximate wavelength range $0.1 \leq \lambda \leq 10$ \um.  As this is the same approximate wavelength range where the emission from the protostar+disk input SEDs peak, neglecting the opacity from scattering will artificially boost the amount of short-wavelength radiation escaping from the models.  Other, more recent studies of individual sources using DUSTY to model the observed SEDs have attempted to correct for this by treating the total opacity ($\kappa_{abs} + \kappa_{scat}$) entirely as absorption (e.g., Bourke et al.~2006; Dunham et al.~2006).  This method will give the correct amount of total opacity, although it will overcorrect and produce too \emph{little} short-wavelength radiation since some of the absorbed radiation should have been scattered instead.

Here we revisit the issue of including scattering in the radiative transfer.  Even with preferential forward-scattering off dust grains, at least some near-infrared emission is still expected.  Indeed, Foster \& Goodman (2006) detect extended near-infrared emission arising from such scattering, which they call ``cloudshine'', in very deep near-infrared images of dark clouds in Perseus.  Unlike near-infrared emission arising from a protostar, which is compact in nature, the cloudshine originates from the full extent of the core.  Thus, assuming typical-sized apertures of a few arcseconds or less are used, only a small amount (approximately equal to the ratio of the solid angle subtended by the aperture to that subtended by the full extent of the core) of this emission would be included in near-infrared photometry of embedded sources.  Furthermore, subtraction of the sky background performed in any standard photometry procedure would remove the small amount of cloudshine that is included in the aperture.  Thus, none of the cloudshine should be included in these models if they are to be compared to observations.

By switching to the 2-D dust radiative transfer package RADMC, we are able to simulate observations by including both small apertures and background subtraction.  Like DUSTY, RADMC assumes that the scattering process is isotropic.  However, as RADMC is a Monte Carlo code that follows individual photons from their creation at the central source through to their final escape from the system, the locations of the source of the observed photons are preserved.  We thus calculate the final SED in apertures of fixed sizes to include the emission from the compact source but exclude most of the diffuse emission from scattering of the ISRF.  Following Crapsi et al.~(2008), we select the aperture radii to approximately match the resolution of the \emph{Spitzer Space Telescope}:  2\as\ (280 AU at 140 pc) for $\lambda \leq 10$ \um, 6\as\ (840 AU at 140 pc) for $10 < \lambda \leq 40$ \um, and 20\as\ (2800 AU at 140 pc) for $40 < \lambda \leq 100$ \um.  Longward of 100 \um\ we use an aperture large enough to encompass the full extent of the model.  To simulate removing any remaining cloudshine through sky background subtraction, we run the entire model grid a second time, including only the ISRF and no internal source of luminosity.  To construct the final SED at each timestep, we then subtract the model with only external heating from the model with both internal and external heating for $\lambda < 10$ \um, since no core heated only externally by the ISRF will emit any significant thermal radiation at such short wavelengths (e.g., Evans et al.~2001).

\begin{figure}[t]
\epsscale{1.0}
\vspace{0.5in}
\plotone{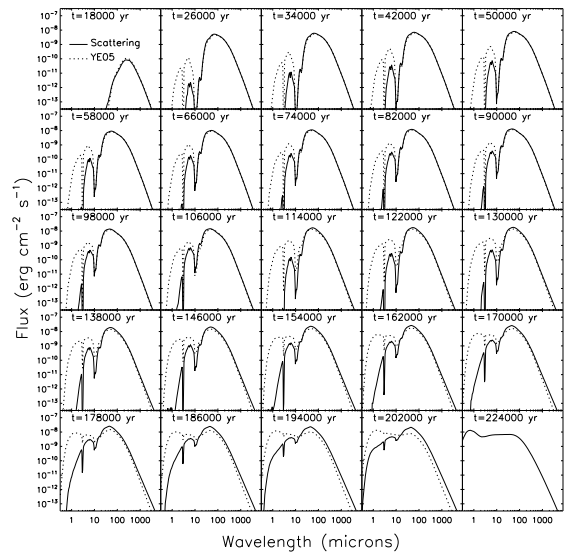}
\caption{\label{fig_mod1_seds}Model 1 spectral energy distributions (SEDs) at select ages through the 224,000 yr collapse of the 1 \msun\ core.  In each panel, the solid line shows the model 1 (\S \ref{mod1}) SED while the dotted line shows the original YE05 SED.  There is no YE05 SED shown in the last panel because the YE05 1 \msun\ model finished collapsing at 210,000 yr.}
\end{figure}

Figure \ref{fig_mod1_seds} compares the SEDs with scattering included to the YE05 SEDs at various times throughout the collapse of the 1 \msun\ core (analogous to Figure 8 in YE05).  As expected, the SEDs with scattering included have significantly less short-wavelength emission than found by YE05.  At later times there is also an increasing discrepancy in the long-wavelength emission.  Since the emission at such wavelengths is optically thin it traces the total envelope mass, signifying a growing difference with time in the total remaining envelope mass.  This is caused by the 5\% lower $\dot{M}_{env}$ in this work compared to the YE05 model (see \S \ref{model_setup}).

\begin{figure}[hbt!]
\epsscale{0.6}
\vspace{0.35in}
\plotone{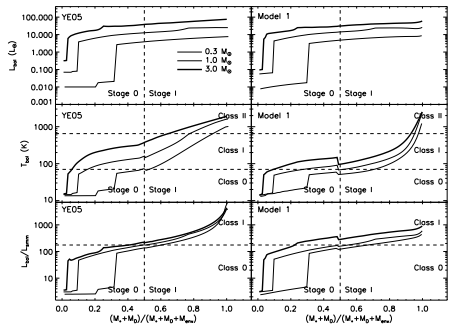}
\caption{\label{fig_mod1_evols1}Observational signatures \lbol\ (top panels), \tbol\ (middle panels), and \lbolsmm\ (bottom panels) versus $M_{int}/M_{tot}$, the ratio of internal (protostar+disk) to total (protostar+disk+envelope) mass.  The left panels show the YE05 model results while the right panels show the model 1 results.  The class boundaries in \tbol\ are taken from Chen et al.~(1995) while the class divisions in \lbolsmm\ are taken from YE05.  The discontinuities in both \tbol\ and \lbolsmm\ are artifacts introduced by the switch from the Shu (1977) density profile to a power-law density profile.  They are present with the same magnitude in the YE05 model but are less readily apparent because of the logarithmic axis scale.}
\end{figure}

Figure \ref{fig_mod1_evols1} shows the evolution of \lbol, \tbol, and \lbolsmm, both for the original YE05 model and for model 1, plotted against the ratio of internal (protostar+disk) to total (protostar+disk+envelope) mass ($M_{int}/M_{tot}$).  Based on the physical definition of Stage 0 as the portion of the embedded phase when at least half of the total system mass is still in the envelope, the model should cross the Class 0/I boundaries in \tbol\ and \lbolsmm\ when $M_{int}/M_{tot} =$ 0.5.  From this, YE05 concluded that \lbolsmm\ is a much more reliable evolutionary indicator than \tbol.  Indeed, Figure \ref{fig_mod1_evols1} shows that, in the YE05 model, the \tbol\ Class 0/I boundary is crossed when $M_{int}/M_{tot} =$ 0.48, 0.15, and 0.05 for the 0.3, 1, and 3 \msun\ cores, respectively.  On the other hand, the YE05 model crosses the \lbolsmm\ Class 0/I boundary when $M_{int}/M_{tot} =$ 0.60, 0.51, and 0.40, respectively.

Including the opacity from scattering changes these results.  As noted above, including scattering significantly decreases the short-wavelength emission since the opacity at these wavelengths is increased by up to a factor of 10 (Figure \ref{fig_scattering}).  As a consequence, the calculated \tbol\ of a given model decreases substantially.  To be quantitative, for the 1 \msun\ core, except for very early times when neither model features short-wavelength emission, the model with scattering included has a calculated \tbol\ lower by a factor of about $1.5-6$.  This reduction in \tbol\ is evident in Figure \ref{fig_mod1_evols1}.  For the model with scattering included, the \tbol\ Class 0/I boundary is crossed when $M_{int}/M_{tot} =$ 0.70, 0.28, and 0.09 for the 0.3, 1, and 3 \msun\ cores, respectively, uniformly later than in the YE05 models.  This same model crosses the \lbolsmm\ Class 0/I boundary when  $M_{int}/M_{tot} =$ 0.66, 0.54, and 0.22, respectively.

Based on these results, we conclude that it is important to include the opacity from scattering at short wavelengths.  Doing so reduces the amount of short-wavelength emission and is thus crucial for comparing model and observed SEDs.  It also reduces the calculated \tbol\ of a given model by a factor of $\sim 1.5-6$ depending on the exact parameters of the model.  It does not significantly affect \lbolsmm, however, since this quantity is much less sensitive to the exact amount of short-wavelength emission (the lower values of \lbolsmm\ at late times compared to the YE05 model is a consequence of the increased long-wavelength emission arising because of the 5\% lower mass accretion rate, as described above, and is unrelated to including the opacity from scattering).  The Class 0/I \tbol\ boundary is still crossed too early for the 1 and 3 \msun\ cores compared to the Stage 0/I boundary, but the discrepancy is not as bad as in the YE05 model.  While \lbolsmm\ remains the most reliable evolutionary indicator for associating physical Stage with observational Class, we caution that, in practice, it is more difficult to calculate and significantly more prone to error than \tbol, depending on the exact submillimeter wavelengths at which observations are available (Dunham et al.~2008).

\begin{figure}[hbt!]
\epsscale{0.90}
\plotone{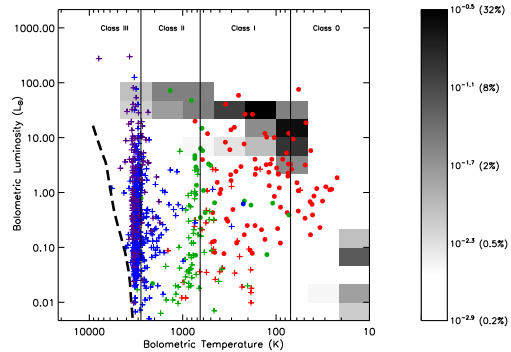}
\caption{\label{fig_mod1_blt}Same as Figure \ref{fig_modye05_blt}, except now for model 1 rather than the YE05 model, and without model tracks.  The grayscale pixels indicate the fraction of total time the model spends in each \lbol\ $-$ \tbol\ bin, calculated from Equation \ref{eq_bins}.  The grayscale is displayed in a logarithmic stretch with the scaling chosen to emphasize the full extent of the models in \lbol\ $-$ \tbol\ space.  The mapping between grayscale and fraction of total time is indicated in the legend.  The class boundaries in \tbol\ are taken from Chen et al.~(1995).  The thick dashed line shows the ZAMS (D'Antona \& Mazzitelli 1994) from 0.1 to 2.0 \msun.  The colored symbols show the Young Stellar Objects from Evans et al.~(2009) in this diagram; the colors and symbols hold the same meaning as in Figure \ref{fig_modye05_blt}.}
\end{figure}

Figure \ref{fig_mod1_blt} shows a BLT diagram for model 1, similar to Figure \ref{fig_modye05_blt} for the YE05 model.  While including the opacity from scattering has important consequences, as described above, the general luminosity problem described in \S \ref{intro} remains.

\begin{figure}[hbt!]
\epsscale{1.0}
\plotone{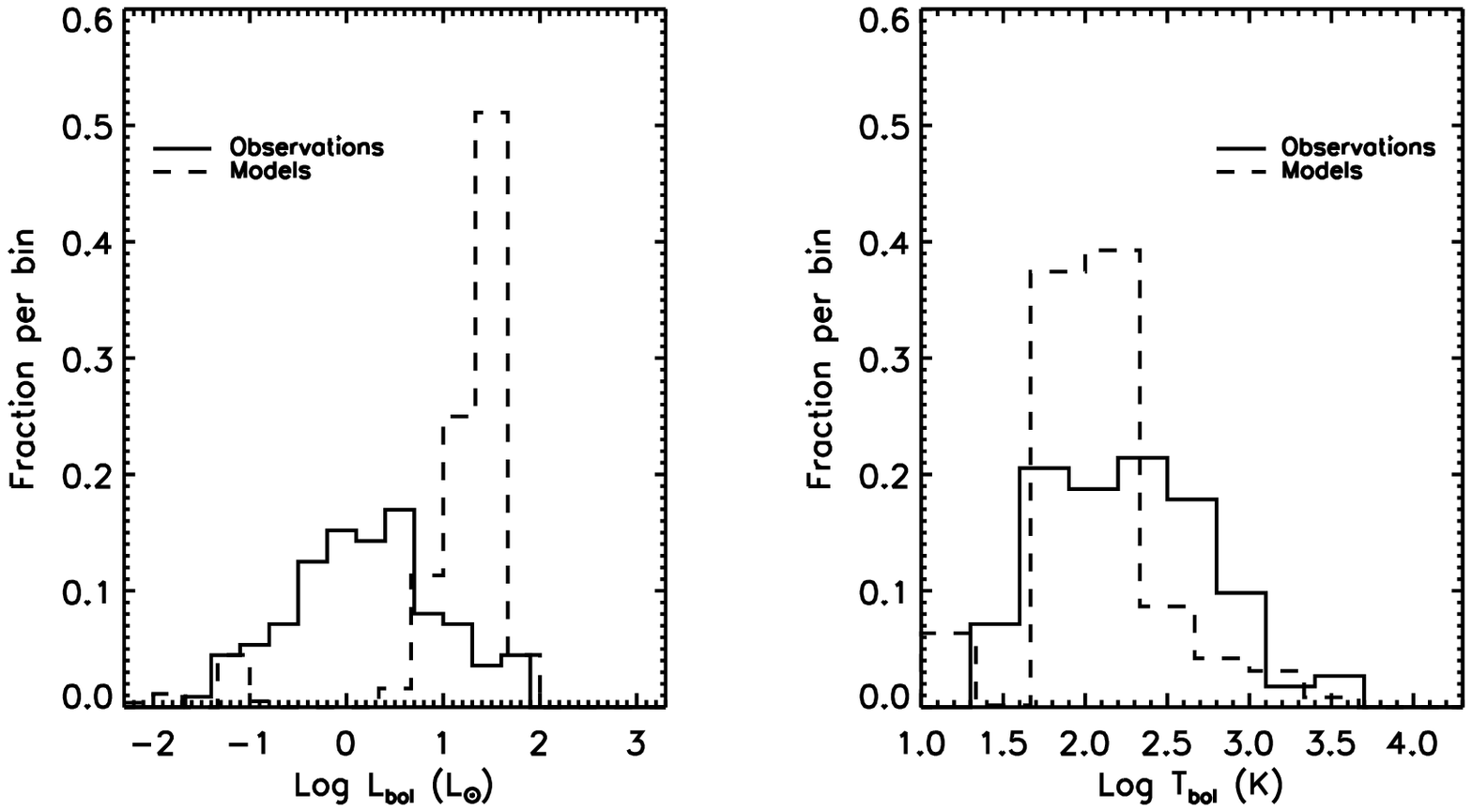}
\caption{\label{fig_mod1_hist}Same as Figure \ref{fig_modye05_hist}, except now for model 1 rather than the YE05 model: histograms showing the fraction of total sources (observations; solid lines) and fraction of total time spent (models; dashed lines; calculated from Equation \ref{eq_bins}) at various \lbol\ (left) and \tbol\ (right).  The binsize is 1/3 dex in both quantities.  For the observations, only the 112 embedded sources (plotted as filled circles on the BLT diagrams) are included.}
\end{figure}

Figure \ref{fig_mod1_hist} shows \lbol\ and \tbol\ histograms for both the observations and model 1, similar to Figure \ref{fig_modye05_hist} for the YE05 model, and column 4 of Table \ref{tab_fractionperbin} lists the fraction of total time model 1 spends in various \lbol\ $-$ \tbol\ bins.  These results again emphasizes the luminosity problem.  76.8\% of the observed sources have $0.1 \leq$ \lbol\ $< 10$ \lsun\ while 16.1\% have $10 \leq$ \lbol\ $< 1000$ \lsun, whereas the models spends only 13.4\% of the time at $0.1 \leq$ \lbol\ $< 10$ \lsun\ but 80.6\% of the time at $10 \leq$ \lbol\ $< 1000$ \lsun.  Furthermore, a K-S test shows that there is less than a 0.1\% probability that the observed and model \lbol\ histograms represent the same underlying distribution.  A similar K-S test gives a 42\% probability that the observed and model \tbol\ histograms represent the same underlying distribution.  The increase in short-wavelength opacity and corresponding decrease in both short-wavelength model emission and model \tbol\ is clearly seen in the \tbol\ histogram in that model 1 spends more time at low \tbol\ (\tbol\ $\la 200$ K) compared to the YE05 model.  Compared to the observations, model 1 overpredicts the fraction of sources observed at these low values of \tbol.

\subsection{Model 2: Two-Dimensional Disk}\label{mod2}

A protostellar disk is inherently a two (or higher) dimensional object, but the radiative transfer in the YE05 model was calculated with DUSTY, a one-dimensional radiative transfer code.  In their model and model 1 above, a disk was included by calculating the emission from a disk with specified parameters, averaging this emission over all inclinations, and then adding the result to the (proto)stellar spectrum to form the final input SED for the radiative transfer calculation (Butner et al.~1994; Adams et al.~1988).  However, since this work is performed with RADMC, a 2-D radiative transfer code, we are able to include a disk directly in the radiative transfer rather than simulate its presence by the above method.

Several other recent authors have published 2-D models of embedded protostars that include disks (e.g., Whitney et al.~2003a; Whitney et al.~2003b; Robitaille et al.~2006; Crapsi et al.~2008).  Following their examples, we include a relatively simple disk with a density profile featuring a power-law in the radial coordinate and a Gaussian in the vertical coordinate:
\begin{equation}\label{eq_disk_density_profile}
\rho_{disk}(s,z) = \rho_0 \left(\frac{s}{s_o} \right)^{-\alpha} exp\left[ -\frac{1}{2} \left(\frac{z}{H_s}\right)^2 \right] \qquad .
\end{equation}
In equation \ref{eq_disk_density_profile}, $z$ is the distance above the midplane ($z=rcos\theta$, with $r$ and $\theta$ the usual radial and zenith angle spherical coordinates) and $s$ is the distance in the midplane from the origin ($s= \sqrt{r^2 - z^2}$).  The quantity $H_s$ is the disk scale height and is given by $H_s = H_o \left(\frac{s}{s_o}\right)^\beta$, where $H_o$ is the scale height at the reference midplane distance $s_o$.  Following Whitney et al.~(2003a), we set $H_o = 10$ AU at $s_o = 100$ AU.  The parameter $\beta$ describes how the scale height changes with $s$ and sets the flaring of the disk, while $\alpha$ describes how the midplane density profile varies with $s$.  Again following Whitney et al.~(2003a), we choose $\beta = 1.25$ and $\alpha = 2.25$.  These values are close to those adopted by Crapsi et al.~(2008; $\beta = 9/7$, $\alpha = 16/7$) to correspond to the self-irradiated passive disk model of Chiang \& Goldreich (1997), and also to the best-fit values found by Sauter et al.~(2009; $\beta = 1.4$, $\alpha = 2.2$) from a detailed model of the edge-on circumstellar disk CB 26.  The mass of the disk is controlled by the parameter $\rho_0$, where the mass, inner disk radius, and outer disk radius evolve following the description in \S \ref{model_setup}.  The disk surface density profile, calculated by integrating Equation \ref{eq_disk_density_profile} over the vertical coordinate $z$, has a radial power-law index of $\Sigma(s) \propto s^{-p}$, where $p = \alpha - \beta$.  With our adopted values of $\alpha$ and $\beta$, $p = 1$.

We include scattering as described in \S \ref{mod1}, and we include the same luminosity components described in \S \ref{model_luminosities}.  In the original model and that presented in \S \ref{mod1}, $L_{EtoD}$, $L_{DtoP}$, and $L_{DM}$ are treated as intrinsic disk luminosity, while $L_{EtoP}$, $L_{DR}$, and $L_{phot}$ are treated as protostellar luminosity.  However, our 2-D radiative transfer package RADMC is limited to one internal source of photons (the protostar) and one external source (the ISRF).  The disk in this model is thus treated as a purely reprocessing disk; it reprocesses radiation from the protostar but does not have its own intrinsic luminosity.  To keep the overall model luminosities correct, we treat all six sources of luminosity as protostellar and calculate the resulting temperature profiles and emergent SEDs of the disk+envelope model.  The main consequence of adopting this purely reprocessing disk is less mid-infrared emission, as discussed in greater detail below.

Before presenting our results, we note that the equations presented by Adams \& Shu (1986) for the luminosity components assume a spatially thin disk (confined to the $z=0$ plane), which is clearly not the case for the flared disk adopted here.  Material accreting onto the disk will not fall all the way to the midplane before joining the disk and thus will not fall as deep into the potential well.  As a consequence, $L_{EtoD}$, calculated assuming a thin disk, is actually an upper limit to the true value.  However, the total model luminosity should be correct, since the decreased luminosity from accretion onto the disk should be compensated by the increased luminosity from the accretion of this material from the disk onto the protostar.  The YE05 model also featured this small physical inconsistency since the parameters of their simulated, 1-D disk were chosen to simulate a flared disk.  The overall effects on the results should be negligible since we are interested in the global evolution of a collapsing protostellar core rather than the details of the individual components of the total luminosity.

\begin{figure}[t]
\epsscale{1.0}
\vspace{0.5in}
\plotone{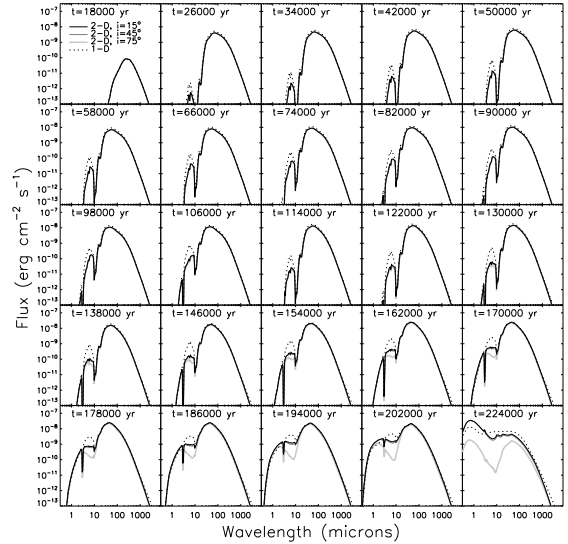}
\caption{\label{fig_mod2_seds}Same as Figure \ref{fig_mod1_seds}, except now for model 2: SEDs at select ages through the collapse of the 1 \msun\ core.  In each panel, the lines show the model 2 SEDs at various inclinations, with the line weight key given in the first panel.  The dotted line shows, for comparison, the SED at that time for model 1.}
\end{figure}

Figure \ref{fig_mod2_seds} compares the model SEDs at various inclinations for the 1 \msun\ core to the model 1 SEDs.  There is very little inclination dependence at early times when the disk is small and not very massive.  At late times, as the disk grows and becomes more massive, the inclination dependence increases, although the only substantial change with inclination occurs for nearly edge-on lines of sight that pass through the disk.

Compared to the SEDs with a 1-D, simulated disk, the main difference of this model (aside from the inclination dependence) is a decrease in the $3-10$ \um\ emission.  This difference arises because model 1 includes intrinsic disk luminosity which raises the disk temperature profile above that arising solely from reprocessing of (proto)stellar radiation, whereas model 2 presented here is limited to a reprocessing disk only.  While the total internal luminosity of the model remains correct, as described above, there is less mid-infrared emission because the disk is generally cooler when heated only by reprocessing.  However, as shown below, this difference has only a small effect on calculated evolutionary indicators and is not important in the context of this work.  The exact amount of mid-infrared emission in the final SEDs can be adjusted by varying the degree of flaring of the disk (through the parameter $\beta$), which will affect the amount of (proto)starlight intercepted and reprocessed by, and thus the temperature profile of, the disk.

\begin{figure}[t]
\epsscale{0.6}
\vspace{0.5in}
\plotone{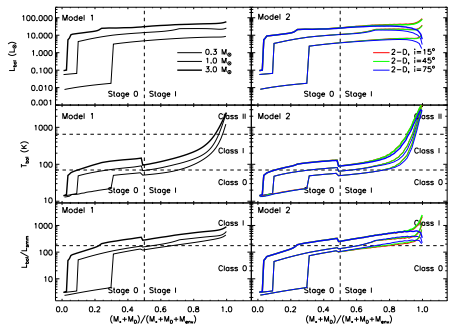}
\caption{\label{fig_mod2_evols1}Same as Figure \ref{fig_mod1_evols1}, except now with model 1 in the left panels and model 2 in the right panels: observational signatures \lbol\ (top panels), \tbol\ (middle panels), and \lbolsmm\ (bottom panels) versus $M_{int}/M_{tot}$, the ratio of internal (protostar+disk) to total (protostar+disk+envelope) mass.  The different colors show the results for different inclinations; the color key is given in the upper right panel.  The class boundaries in \tbol\ are taken from Chen et al.~(1995) while the class divisions in \lbolsmm\ are taken from YE05.  The discontinuities in both \tbol\ and \lbolsmm\ are artifacts introduced by the switch from the Shu (1977) density profile to a power-law density profile.}
\end{figure}

Figure \ref{fig_mod2_evols1} shows the observational signatures \lbol, \tbol, and \lbolsmm\ plotted against the ratio of $M_{int}/M_{tot}$, with different colors corresponding to different inclinations.  As expected from the SEDs, there is essentially no inclination dependence visible in any quantity except at late times, and even then most of the dependence is seen in the nearly edge-on line-of-sight ($i=75$\degree) relative to the other lines of sight.  To be quantitative, for the 1 \msun\ core, \tbol\ calculated from SEDs at the same time viewed at 5\degree\ and 85\degree\ vary by $<5$\% for all times up to $\sim$150,000 yr, and by $<25$\% for the remaining times.  The ratio of \lbolsmm\ shows similar results: \lbolsmm\ calculated from SEDs viewed at 5\degree\ and 85\degree\ varies by $<25$\% except at very late times (approximately the last $10^4$ yr), where the nearly edge-on lines-of-sight passes through a disk that has become so optically thick that the calculated \lbol\ significantly decreases (evident in the last panel of Figure \ref{fig_mod2_seds}).  At these late times, \lbolsmm\ changes by about a factor of four from pole-on to edge-on inclinations.

Compared to model 1, the addition of a 2-D disk in the radiative transfer causes small reductions in both \tbol\ and \lbolsmm.  This is caused by the change to a reprocessing disk described above, which shifts some of the shorter-wavelength, $3-10$ \um\ emission to longer wavelengths and thus decreases both evolutionary indicators.  The effect is small; at any given time, \tbol\ decreases by $<25$\% and \lbolsmm\ decreases by $<15$\%.  As a consequence, model 2 crosses the Class 0/I boundary\footnote{Different inclinations cross the Class 0/I boundary at slightly different times, and thus at slightly different values of $M_{int}/M_{tot}$.  The values presented here are calculated by taking a weighted average of the values at each inclination, with the inclination weights as defined in the discussion following Equation \ref{eq_bins} (\S \ref{mod1}).} slightly later:  the \tbol\ Class 0/I boundary is crossed when $M_{int}/M_{tot} =$ 0.76, 0.31, and 0.11 for the 0.3, 1, and 3 \msun\ cores, respectively, and the \lbolsmm\ Class 0/I boundary is crossed when $M_{int}/M_{tot} =$ 0.74, 0.61, and 0.23, respectively.  These times are slightly later than model 1 (see \S \ref{mod1}), but the overall conclusions about the connection between physical Stage and observational Class, as defined by \tbol\ and \lbolsmm, remain unchanged.

A careful inspection of Figure \ref{fig_mod2_evols1} reveals small-scale oscillations in both \tbol\ and \lbolsmm\ at late times.  These are artifacts introduced by the model gridding.  The grid is logarithmically spaced in radius to ensure that there are enough gridpoints at small radii where the optical depth is large.  As a consequence, the spacing between gridpoints becomes relatively large ($\sim 5-10$ AU per gridpoint) at $50-100$ AU, the range of disk outer radii at these late times.  As the disk radius grows with time, it generally takes a few timesteps before it increases enough to ``jump'' to the next gridpoint.  In the model it thus remains fixed at the previous gridpoint, but as the mass is increasing at each timestep, the optical depth through the disk is also increasing.  Once the radius increases enough to jump to the next gridpoint, the optical depth suddenly decreases slightly, affecting the temperature profile, emergent SEDs, and thus calculated evolutionary indicators.  These effects are small enough to have a negligible impact on the results.

\begin{figure}[hbt!]
\epsscale{0.90}
\plotone{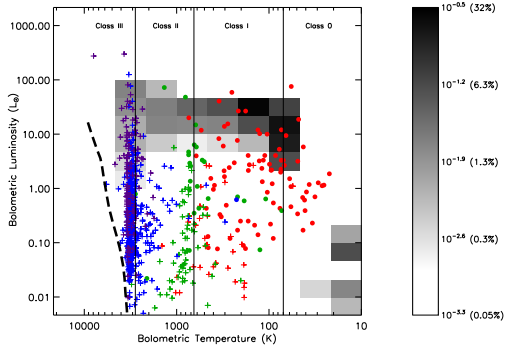}
\caption{\label{fig_mod2_blt}Same as Figure \ref{fig_mod1_blt}, except now for model 2.  The grayscale pixels indicate the fraction of total time the model spends in each \lbol\ $-$ \tbol\ bin, calculated from Equation \ref{eq_bins}.  The grayscale is displayed in a logarithmic stretch with the scaling chosen to emphasize the full extent of the models in \lbol\ $-$ \tbol\ space.  The mapping between grayscale and fraction of total time is indicated in the legend.  The class boundaries in \tbol\ are taken from Chen et al.~(1995).  The thick dashed line shows the ZAMS (D'Antona \& Mazzitelli 1994) from 0.1 to 2.0 \msun.  The colored symbols show the Young Stellar Objects from Evans et al.~(2009) in this diagram; the colors and symbols hold the same meaning as in Figure \ref{fig_mod1_blt}.}
\end{figure}

\begin{figure}[hbt!]
\epsscale{1.0}
\plotone{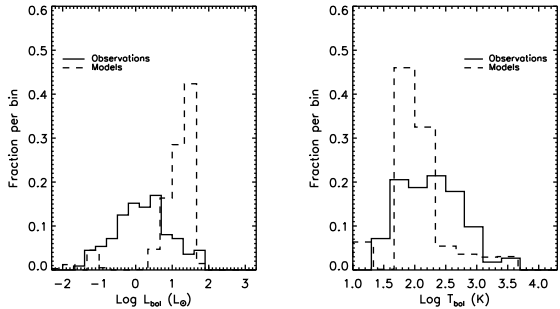}
\caption{\label{fig_mod2_hist}Same as Figure \ref{fig_mod1_hist}, except now for model 2: histograms showing the fraction of total sources (observations; solid lines) and fraction of total time spent (models; dashed lines; calculated from Equation \ref{eq_bins}) at various \lbol\ (left) and \tbol\ (right).  The binsize is 1/3 dex in both quantities.  For the observations, only the 112 embedded sources (plotted as filled circles on the BLT diagrams) are included.}
\end{figure}

Figures \ref{fig_mod2_blt} and \ref{fig_mod2_hist} show a BLT diagram and \lbol\ and \tbol\ histograms for model 2, respectively, and column 5 of Table \ref{tab_fractionperbin} lists the fraction of total time model 2 spends in various \lbol\ $-$ \tbol\ bins.  The inclination dependence introduced by the disk results in slightly more spread in model coverage in \lbol\ $-$ \tbol\ space, and shifts the peak of the model luminosity distribution to slightly lower luminosities, but the main model 1 conclusions that the model overpredicts both the time spent at high luminosities ($\ga 1$ \lsun) and the time spent at low \tbol\ ($\la 200$ K) remain unchanged.  Indeed, K-S tests on Figure \ref{fig_mod2_hist} give similar results as those performed on Figure \ref{fig_mod1_hist}:  there is again less than a 0.1\% probability that the observed and model \lbol\ histograms represent the same underlying distribution, and a 34\% probability of the same for the \tbol\ histograms.

\subsection{Model 3: Two-Dimensional Envelope}\label{mod3}

All models presented up to this point have featured the spherically symmetric density profiles calculated by Shu (1977) for the collapse of singular isothermal spheres.  In reality, a collapsing core that is initially spherically symmetric will not remain so if the core has any initial rotation; conservation of angular momentum will create a rotationally flattened structure.  As the models presented here do feature initial rotation (set by the initial angular velocity $\Omega_0$; see \S \ref{model_setup}), this effect should be included.  The move to the 2-D radiative transfer code RADMC enables us to do this.

To include the effects of rotation on the evolution of the model, we adopt the solution for the collapse of a slowly rotating core presented by TSC84.  The core is initially a spherically symmetric, singular isothermal sphere with a density distribution $n \propto r_{env}^{-2}$, identical to the Shu (1977) solution.  The outer radius is again truncated to set the initial core mass.  As collapse proceeds, the solution takes on two forms: an outer solution that is similar to the non-rotating, spherically symmetric solution and an inner solution that exhibits flattening of the density profile.  Since material falling in to the central regions originates from larger radii and thus carries more angular momentum as time progresses, the radius where the inner solution must be used, and thus the radius at which flattening becomes significant, increases with time ($r_{flat} \propto \Omega_0^2 t^3$; TSC84), reaching $\sim$ 2000 AU at late times for all three initial mass cores (the longer collapse times of the higher mass cores are offset by slower initial angular velocities).

\begin{figure}[hbt!]
\epsscale{1.0}
\plotone{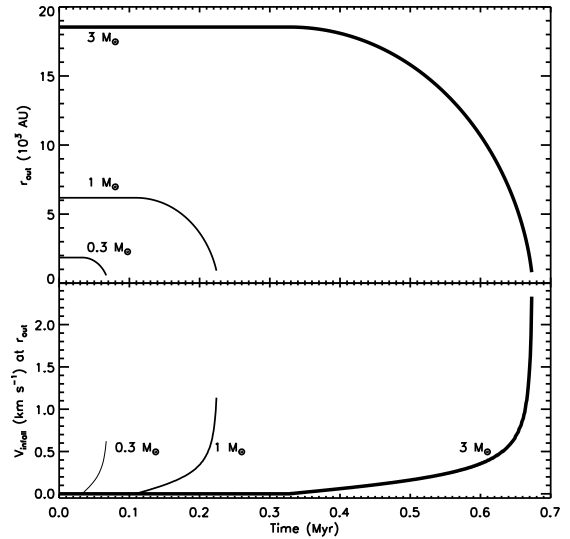}
\caption{\label{fig_mod3_rout}The model 3 envelope outer radius (top panel) and inward radial velocity at this outer radius (bottom panel) as a function of time for the 0.3, 1, and 3 \msun\ cores.  The outer radius of each core remains constant with no inward radial velocity for the first half of the collapse when the infall radius has not yet reached the outer boundary of the core, and decreases with an increasing inward radial velocity once the infall radius reaches the outer boundary.}
\end{figure}

Once half of the initial core mass has accreted onto the protostar+disk system, the infall radius, which moves outward at the sound speed, exceeds the envelope outer radius.  If we simply continued to use the TSC84 solution, ``extra mass'' that originated beyond the outer radius would begin to collapse and eventually move within this radius.  Thus, there would still be mass remaining in the envelope once the initial mass of the core has accreted, which is clearly not self-consistent (although see Myers [2008] for an analytic model that includes protostellar accretion from a core embedded in a uniform background that also partially accretes onto the protostar).  YE05 also faced this problem in their 1-D model and avoided it by simply switching to an $n \propto r_{env}^{-3/2}$ power-law density profile once the infall radius exceeded the envelope outer radius, with the outer radius kept fixed and the desired mass at each timestep in the model evolution achieved by adjusting the normalization of the power-law.  However, this results in a model that is no longer physically self-consistent, since it does not make sense to have a core that is collapsing at all radii (which is the case once the infall radius exceeds the outer radius) and thus decreasing in mass but remaining a fixed size.  Furthermore, such a solution is not available to us here since we wish to retain the feature of the TSC84 solution that the radius at which the density profile exhibits flattening increases with time.  Thus, as an alternative, we use the velocity profiles given by the TSC84 solution and allow the outer radius to decrease once the infall radius exceeds the initial outer radius.  This is illustrated by Figure \ref{fig_mod3_rout}, which shows the outer radius of the core and inward radial velocity at this outer radius as a function of time for the three different initial mass cores.  The effect of this change in the calculation of the density profiles for the second half of the collapse of each core is discussed below.

Figure \ref{fig_mod3_seds} compares the model 3 SEDs at various inclinations for the 1 \msun\ core to the model 1 SEDs presented in \S \ref{mod1}.  As was the case for model 2, the inclination dependence is small at early times and increases throughout the collapse of the core.  SEDs at late times show noticeably less short-wavelength emission.  Part of this is due to the change to a purely reprocessing disk, as discussed above.  However, this deficit in short-wavelength emission becomes especially noticeable at very late times and has a second cause: the decrease in the outer radius of the envelope as material at the initial outer edge collapses.  The reason for this is best understood by assuming this is a 1-D, spherically symmetric model with a power-law density profile given by $\rho(r) = \rho_f (r/r_f)^{-p}$, with $\rho_f$ the density at a fiducial radius $r_f$.  Assuming that $1 < p < 3$, the equations for envelope mass and optical depth are:

\begin{figure}[hbt!]
\epsscale{1.0}
\vspace{0.5in}
\plotone{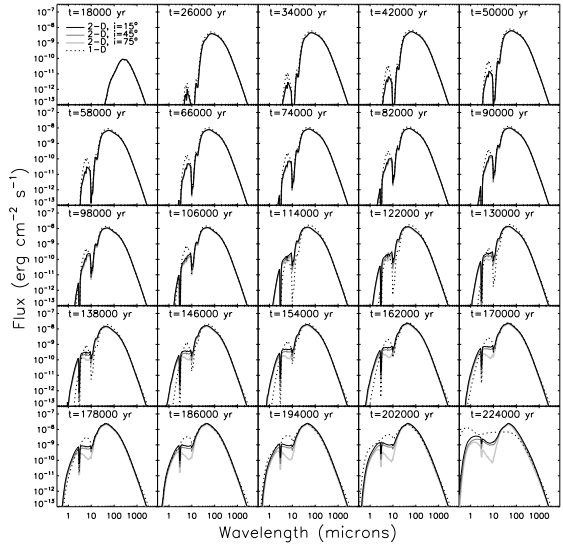}
\caption{\label{fig_mod3_seds}Same as Figure \ref{fig_mod2_seds}, except now for model 3: SEDs at select ages through the collapse of the 1 \msun\ core.  In each panel, the lines show the model 3 (scattering, disk, and rotationally flattened envelope included) SEDs at various inclinations, with the line weight key given in the first panel.  The dotted line shows, for comparison, the SED at that time for model 1.}
\end{figure}

\begin{equation}\label{eq_menv_example}
M_{env} = \int_{r_{in}}^{r_{out}} 4 \pi r^2 \rho(r) dr \sim \frac{4 \pi \rho_f r_f^p}{3-p} r_{out}^{3-p}
\end{equation}
\begin{equation}\label{eq_tau_example}
\tau_{\nu} = \int_{r_{in}}^{r_{out}} \kappa_{\nu} \rho(r) dr \sim \frac{\kappa_{\nu} \rho_f r_f^p}{p-1} r_{in}^{1-p}
\end{equation}

In comparison to model 1 at the same timestep, the envelope has a smaller outer radius but identical mass, which, according to Equation \ref{eq_menv_example}, requires a larger value of $\rho_f$, the normalization of the power-law.  According to Equation \ref{eq_tau_example}, since the envelope inner radius is unchanged, this increases the optical depth through the envelope, giving less short-wavelength emission.  The situation is slightly more complicated in reality since we are comparing a 1-D, spherically symmetric model to a 2-D, rotationally flattened model, but the analogy holds.  Allowing the outer radius to decrease following the TSC84 collapse solution increases the optical depth compared to YE05 and models $1-2$ at the same timestep, where the outer radius is held fixed, and causes more of the short-wavelength emission to be reprocessed to longer wavelengths than for these previous models.

\begin{figure}[hbt!]
\epsscale{0.6}
\vspace{0.35in}
\plotone{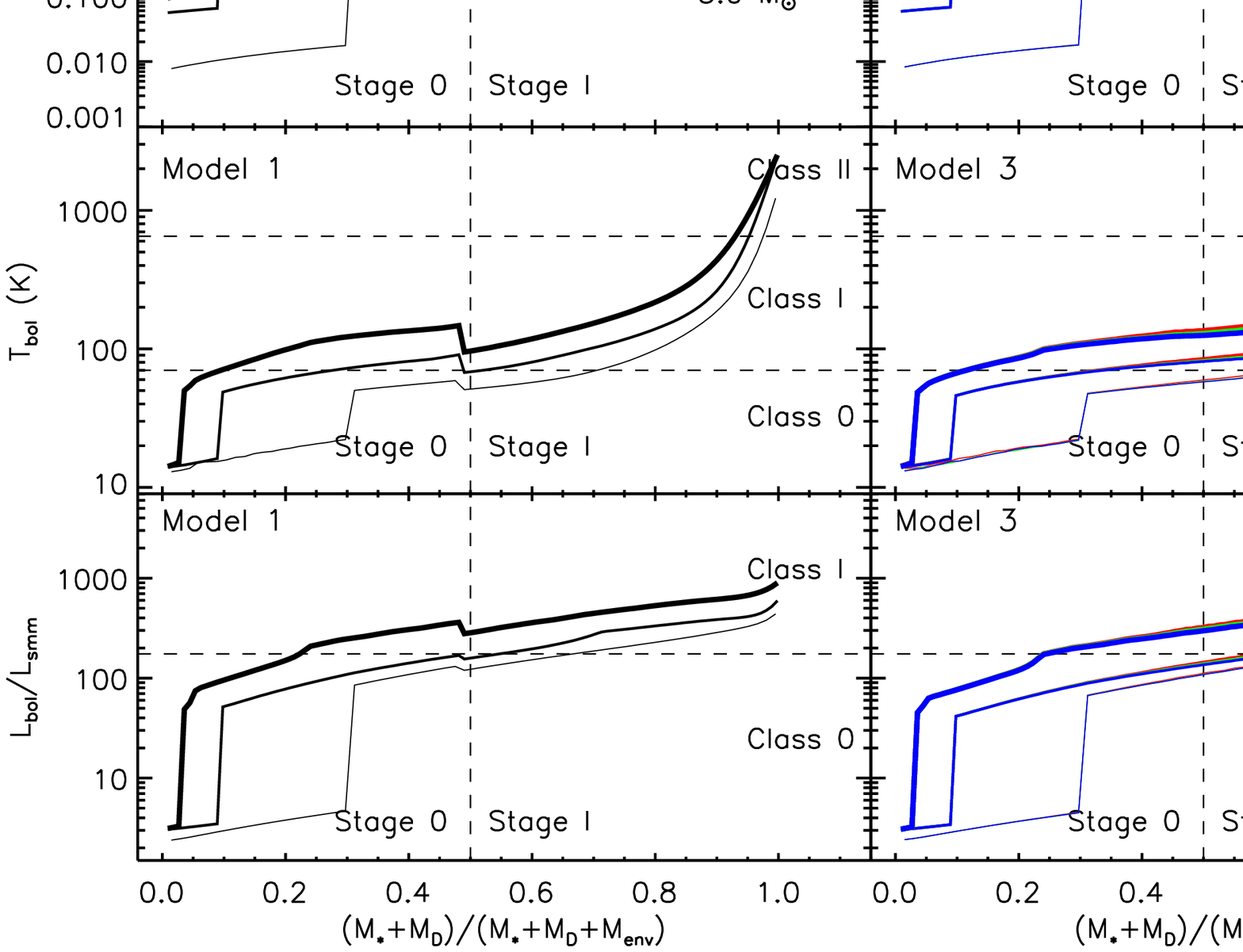}
\caption{\label{fig_mod3_evols1}Same as Figure \ref{fig_mod2_evols1}, except now for model 3: observational signatures \lbol\ (top panels), \tbol\ (middle panels), and \lbolsmm\ (bottom panels) versus $M_{int}/M_{tot}$, the ratio of internal (protostar+disk) to total (protostar+disk+envelope) mass.  The left panels show the model 1 results while the right panels show the model 3 results.  The different colors show the results for different inclinations; the color key is given in the upper right panel.  The class boundaries in \tbol\ are taken from Chen et al.~(1995) while the class divisions in \lbolsmm\ are taken from YE05.  The discontinuities in both \tbol\ and \lbolsmm\ for model 1 are artifacts introduced by the switch from the Shu (1977) density profile to a power-law density profile.}
\end{figure}

Figure \ref{fig_mod3_evols1} shows the observational signatures \lbol, \tbol, and \lbolsmm\ plotted against the ratio of $M_{int}/M_{tot}$, with different colors corresponding to different inclinations.  Although neither \tbol\ nor \lbolsmm\ increases to values as large as those reached in model 2 at late times due to the decrease in short-wavelength emission described above, most of the evolution is similar to model 2.  The \tbol\ Class 0/I boundary is crossed\footnote{Again calculated as a weighted average of the different inclinations, as described in \S \ref{mod2}} when $M_{int}/M_{tot} =$ 0.68, 0.33, and 0.12 for the 0.3, 1, and 3 \msun\ cores, respectively, and the \lbolsmm\ Class 0/I boundary is crossed when $M_{int}/M_{tot} =$ 0.73, 0.61, and 0.24, respectively.  Comparison to the model 2 results shows that the overall conclusions about the connection between physical Stage and observational Class, as defined by \tbol\ and \lbolsmm, again remain unchanged.

Both quantities show an inclination dependence that increases with time.  Using the 1 \msun\ core as an example, \tbol\ calculated from SEDs viewed at 5\degree\ and 85\degree\ vary by $<5$\% only for the first 96,000 yr (compared to the first 150,000 yr for model 2), and the variation increases to 55\% for very late times in the model evolution (compared to 25\% for model 2).  This inclination dependence is not confined solely to lines-of-sight viewed close to edge-on since the flattening of the envelope affects all viewing angles: \tbol\ calculated from SEDs viewed at 5\degree\ and 25\degree\ vary by up to 20\% at late times, and \tbol\ calculated from SEDs viewed at 5\degree\ and 45\degree\ vary by up to 40\% at late times.  For comparison, model 2 showed $< 5$\% variations in \tbol\ at any given time for all inclinations $\la 70$\degree.  Similar results are found for the ratio of bolometric to submillimeter luminosity.  \lbolsmm\ calculated from SEDs viewed at 5\degree\ and 85\degree\ vary by up to 50\% at late times, compared to 25\% (except for the last 10,000 years) for model 2.

While the inclusion of a disk and rotationally flattened envelope following TSC84 does induce a moderate inclination dependence, we note here that it is a much smaller dependence than found by other authors investigating 2-D models of embedded protostars (e.g., Whitney et al.~2003a; Whitney et al.~2003b; Robitaille et al.~2006; Crapsi et al.~2008).  For example, Crapsi et al.~(2008) found a variation in \tbol\ with inclination ranging from factors of $\sim 2-5$ depending on the exact model parameters.  Most of the explanation for this difference resides in the fact that we do not yet include outflow cavities, whereas other authors do.  This will be addressed in \S \ref{mod4}.  However, there is a second, important difference in our models that is worth pointing out:  we follow the exact TSC84 solution whereas these other studies do not.  Instead, they adopt density profiles that exhibit rotational flattening \emph{at all radii}.  As noted by Terebey et al.~(2006), these other models only agree with the TSC84 solution at small radii where the inner solution is valid; at large radii they diverge.  

\begin{figure}[hbt!]
\epsscale{0.90}
\plotone{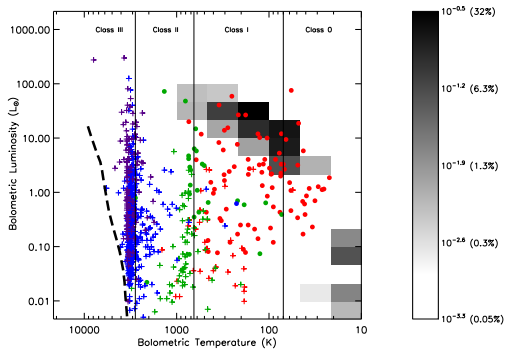}
\caption{\label{fig_mod3_blt}Same as Figure \ref{fig_mod2_blt}, except now for model 3.  The grayscale pixels indicate the fraction of total time the model spends in each \lbol\ $-$ \tbol\ bin, calculated from Equation \ref{eq_bins}.  The grayscale is displayed in a logarithmic stretch with the scaling chosen to emphasize the full extent of the models in \lbol\ $-$ \tbol\ space.  The mapping between grayscale and fraction of total time is indicated in the legend.  The class boundaries in \tbol\ are taken from Chen et al.~(1995).  The thick dashed line shows the ZAMS (D'Antona \& Mazzitelli 1994) from 0.1 to 2.0 \msun.  The colored symbols show the Young Stellar Objects from Evans et al.~(2009) in this diagram; the colors and symbols hold the same meaning as in Figure \ref{fig_mod1_blt}.}
\end{figure}

\begin{figure}[hbt!]
\epsscale{1.0}
\plotone{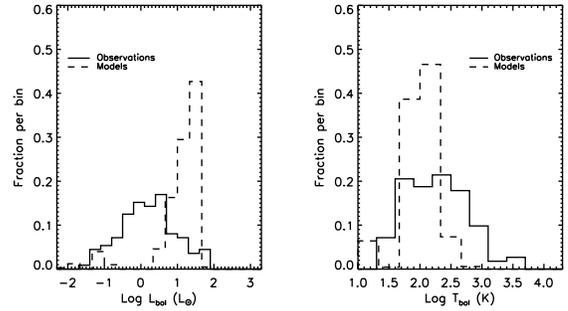}
\caption{\label{fig_mod3_hist}Same as Figure \ref{fig_mod2_hist}, except now for model 3: histograms showing the fraction of total sources (observations; solid lines) and fraction of total time spent (models; dashed lines; calculated from Equation \ref{eq_bins}) at various \lbol\ (left) and \tbol\ (right).  The binsize is 1/3 dex in both quantities.  For the observations, only the 112 embedded sources (plotted as filled circles on the BLT diagrams) are included.}
\end{figure}

Figures \ref{fig_mod3_blt} and \ref{fig_mod3_hist} show a BLT diagram and \lbol\ and \tbol\ histograms for model 3, and column 6 of Table \ref{tab_fractionperbin} lists the fraction of total time model 3 spends in various \lbol\ $-$ \tbol\ bins.  The decrease in \tbol\ at late times due to allowing the outer radius of the core to shrink as material initially at this outer radius collapses is seen in Figure \ref{fig_mod3_blt} in that the models do not extend beyond about 1000 K, and in Figure \ref{fig_mod3_hist} in the increased amount of time spent at \tbol\ $\la 200$ K relative to that spent at \tbol\ $\ga 200$ K.  However, the luminosity distribution remains essentially unchanged, and the main model 1 and model 2 conclusions that the model overpredicts both the time spent at high luminosities ($\ga 1$ \lsun) and the time spent at low \tbol\ ($\la 200$ K) remain unchanged.  Indeed, K-S tests on Figure \ref{fig_mod3_hist} give similar results as those performed on Figure \ref{fig_mod2_hist}:  there is again less than a 0.1\% probability that the observed and model \lbol\ histograms represent the same underlying distribution, and a 13\% probability for the \tbol\ histograms.  The lower probability that the \tbol\ distributions are the same compared to models $1-2$ (13\% versus $\sim 35-40$\%) arises from the shift to even more time spent at low \tbol\ in the model compared to observations.

\subsection{Model 4: Mass-Loss and Outflow Cavities}\label{mod4}

All of the models considered so far have featured a 100\% collapse efficiency\footnote{We refrain from using the term star formation efficiency here, as this term is more commonly used to refer to the fraction of mass initially in the core that ends up in the star.  By that definition, the YE05 model and models 1-3 in this work all feature star formation efficiencies of 75\% (set by the parameter $\eta_D$; see \S \ref{model_setup}).}; all material initially in the core is in the star+disk system at the end of collapse.  In reality, however, this is not the case.  Some of the mass is instead ejected from the system, entraining and removing envelope material as it propagates through the core (e.g., Bontemps et al.~1996; Bally et al.~2007; Arce et al.~2007).  Here we incorporate a simple, idealized process of mass-loss and the opening of outflow cavities into the evolutionary model to test whether or not the increased inclination dependence introduced by outflow cavities can add enough spread to calculated bolometric luminosities and temperatures to bring the model in agreement with observations without resorting to episodic accretion.

A fraction of the material accreted from the disk to the protostar is ejected in the form of a jet or wind.  The exact value of this fraction, $\dot{M}_J / \dot{M}_{acc}$, depends on uncertain models of how the jet/wind is launched.  As compiled by Bontemps et al.~(1996), possible values are $0.1-0.5$ (Shu et al.~1994), $\sim 0.1$ (Pelletier \& Pudritz 1992), and $10^{-5}-0.1$ (Wardle \& K\"{o}nigl 1993).  More recently, observations of the variation of jet diameters with distance from their driving sources have been consistent with models giving $\dot{M}_J / \dot{M}_{acc} > 0.03$ (Ray et al.~2007 and references therein).  Given the above range, we assume $\dot{M}_J / \dot{M}_{acc} = 0.1$; 10\% of the mass accreting from the disk to the protostar is instead ejected from the system in a jet/wind, decreasing the growth of the protostellar mass accordingly.

The jet/wind entrains and removes material as it propagates through the envelope, driving a molecular outflow (e.g., Bontemps et al.~1996; Bally et al.~2007; Arce et al.~2007; although see Machida et al.~2008 for an alternate explanation of the origin of molecular outflows).  Conservation of momentum gives

\begin{equation}\label{eq_consmom}
f \dot{M}_J v_J = \dot{M}_o v_o \qquad ,
\end{equation}
where $\dot{M}_J$ and $\dot{M}_o$ are the mass-loss rates of the jet/wind and outflow, respectively, $v_J$ and $v_o$ are the jet/wind and outflow velocities, respectively, and $f$ is the efficiency with which the jet/wind transfers its momentum to the ambient medium.  We assume a typical jet velocity of 150 km s$^{-1}$ (Bontemps et al.~1996; Bally et al.~2007), consistent with being greater than the $6-60$ km s$^{-1}$ escape velocities of $0.1-1$ \msun\ protostars assuming jet launching radii of $0.5-5$ AU (Ray et al.~2007).  We assume an outflow velocity of 10 km s$^{-1}$, consistent with observations of outflows with typical velocities of $4-5$ km s$^{-1}$ for low-luminosity Class 0 sources (M. Dunham et al.~2009, in preparation; J.E. Lee et al.~2009, in preparation), $5-15$ km s$^{-1}$ for embedded sources in Perseus (Hatchell et al.~2007, 2009), and $\sim$10 km s$^{-1}$ for the sample of 45 embedded sources compiled by Bontemps et al.~(1996).  The efficiency with which the jet/wind transfers its momentum to the ambient medium is not well characterized and likely varies with environment (e.g., Moraghan et al.~2008).  We assume an efficiency of $f=1$ to maximize the effects of mass loss and opening of outflow cavities on the results.

Given the above assumptions, the amount of envelope mass entrained by the jet/wind is $M_{entrained} = 15 M_{ejected}$.  Thus, 10\% of the mass accreting from the disk to the protostar is instead ejected, and 15 times the mass of this ejected material is entrained in the outflow.  The removal of the entrained material is implemented into the model by opening outflow cavities (assumed to be completely devoid of material) in the envelope.  Following Crapsi et al.~(2008), we assume that the outflow cavities follow streamlines of the collapse solution, giving funnel-shaped cavities that are conical at large radii.  The size of the outflow cavity, defined by $\theta_c$, the semi-opening angle of the cone at large radii (see Crapsi et al.~[2008] for details), increases with time to remove the mass entrained at each timestep.  Assuming spherical symmetry of the envelope\footnote{While the envelope density profiles used here are the TSC84 profiles and are thus not spherically symmetric (see \S \ref{mod3}), spherical symmetry is still a good approximation at large radii where most of the mass is located.}, the ratio of mass entrained and removed to the total envelope mass is given by the ratio of the solid angle of the cavity opened to the total solid angle of the envelope, which, assuming a pre-existing cavity with size $\theta_{c,old}$ is already present, is given as:

\begin{displaymath}
\frac{M_{entrained}}{M_{total}} = 2 \, \frac{\Omega_{cavity}}{\Omega_{total}} = 2 \, \frac{\int_0^{2\pi} \int_{\theta_{c,old}}^{\theta_{c,new}} sin\,\theta \: d\theta \: d\phi}{\int_0^{2\pi} \int_{\theta_{c,old}}^{\pi} sin\,\theta \: d\theta \: d\phi} =
\end{displaymath}
\begin{equation}\label{eq_cavityangle}
2 \, \frac{(cos \,\theta_{c,old} - cos \,\theta_{c,new})}{(1 + cos \,\theta_{c,old})} \qquad .
\end{equation}

The factor of 2 accounts for the bipolar nature of the cavities.  The semi-opening angle of the outflow cavity, $\theta_{c,new}$, is calculated from Equation \ref{eq_cavityangle} at each timestep, with the entrained mass at that timestep calculated as described above and $M_{total}$ the total envelope mass at that timestep.

The opening of outflow cavities causes a decrease in $\dot{M}_{env}$, the rate at which material is accreting from the envelope onto the protostar+disk system, since material is no longer accreting from the full $4\pi$ steradian.  As above, the mass accretion rate is calculated from the ratio of solid angles as follows:

\begin{equation}\label{eq_mdotnew}
\frac{\dot{M}_{env}^{new}}{\dot{M}_{env}^{orig}} = 1 - 2 \, \frac{\Omega_{cavity}}{\Omega_{total}} = 1 - 2 \, \frac{\int_0^{2\pi} \int_{0}^{\theta_{c}} sin\,\theta \: d\theta \: d\phi}{\int_0^{2\pi} \int_{0}^{\pi} sin\,\theta \: d\theta \: d\phi} = cos \,\theta_{c} \qquad ,
\end{equation}
where the factor of 2 again accounts for the bipolar nature of the cavities and $\dot{M}_{env}^{orig} = 4.57 \times$ 10$^{-6}$ \msun\ yr$^{-1}$.  We neglect the effects of opening cavities and removing mass on the collapse solution itself; the overall effect would be to slow the collapse of the core (see Myers 2008).

\begin{figure}[t]
\epsscale{0.6}
\vspace{0.5in}
\plotone{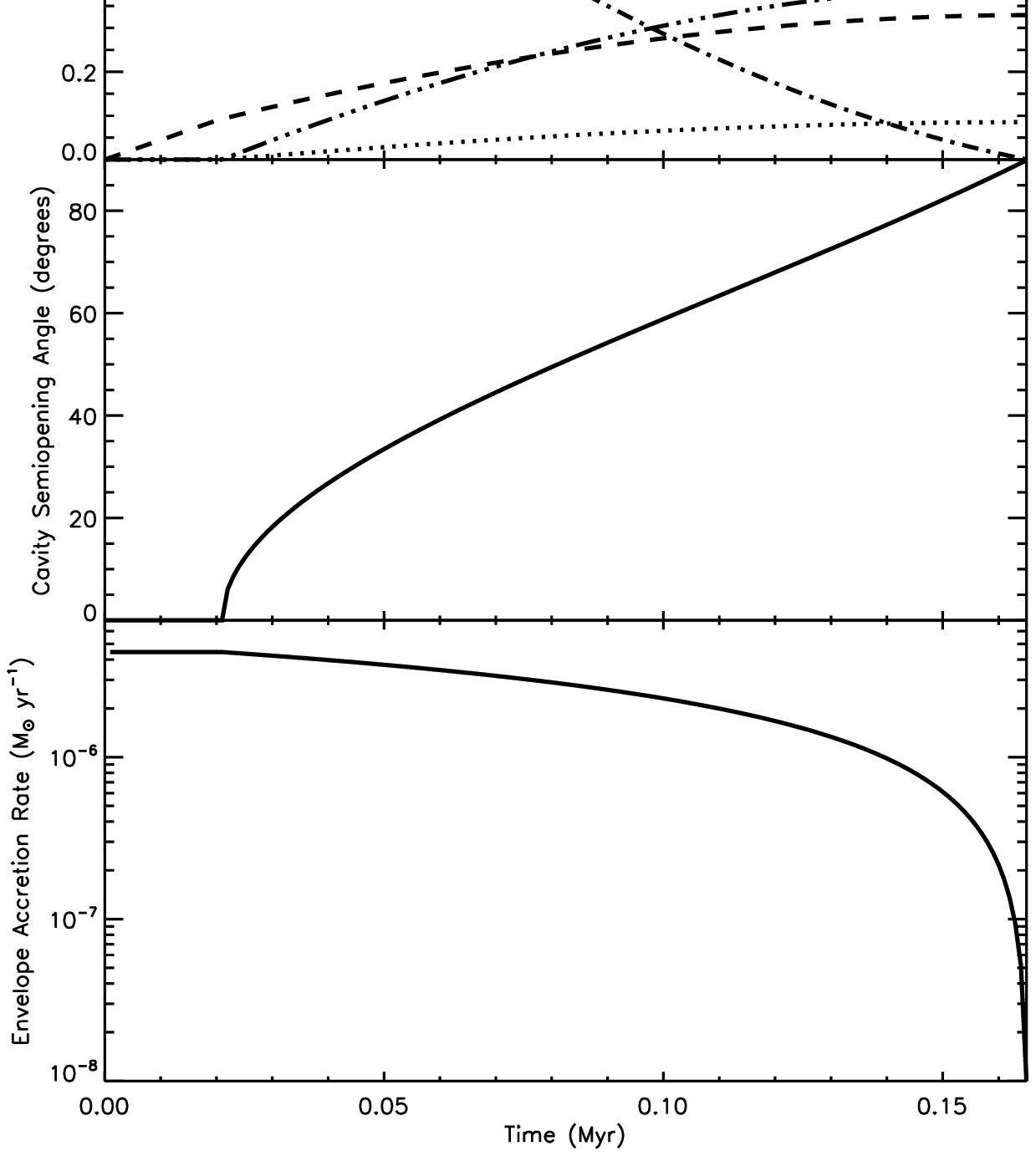}
\caption{\label{fig_mod4_setup}Evolution of the masses of various model components (top panel), outflow cavity semi-opening angle (middle panel), and envelope mass accretion rate (bottom panel) versus time for the model 4 1 \msun\ initial mass core, which includes mass-loss and outflow cavities as described in \S \ref{mod4}.}
\end{figure}

Figure \ref{fig_mod4_setup} shows the effects of including mass loss and outflow cavities as described above for the 1 \msun\ initial mass core.  Plotted are the masses of the various model components (protostar, disk, envelope, and outflow), $\theta_c$, and $\dot{M}_{env}$ versus time.  The envelope mass decreases with time, with a change in slope to a faster decrease once the disk forms and mass is ejected.  At this stage the outflow mass begins to grow, reaching 0.4 \msun\ by the end of collapse.  The envelope mass accretion rate decreases as the outflow cavity increases in size.  Eventually $\theta_c$ reaches 90\degree\ and the collapse ends at 165,000 yr.  Compared to the 224,000 yr collapse duration when no mass-loss is included, this model is $\sim$25\% shorter in total duration.  Similar ($\sim20-30$\%) decreases in collapse duration are seen in the 0.3 and 3 \msun\ initial mass cores.

Defining the dense core star formation efficiency, $f_{SFE}$, as the fraction of material initially in the core that ends up in the star at the end of collapse, this model has $f_{SFE} = 0.48, 0.33,$ and $0.31$ for the 0.3, 1, and 3 \msun\ initial mass cores.  The higher $f_{SFE}$ for the 0.3 \msun\ initial mass core comes about because of the higher fraction of total collapse time spent in the FHSC phase ($\sim$ 40\%, compared to $5-10$\% for the 1 and 3 \msun\ initial mass cores) where there is no mass-loss.  These values of $f_{SFE}$ are in general agreement with those found by Alves et al.~(2007; 30\%) by comparing the CMF of dense cores in the Pipe Nebula to the stellar IMF and by Enoch et al.~(2008; $> 25\%$) by comparing the CMF of dense cores in Perseus, Ophiuchus, and Serpens to the stellar IMF.  Although the modifications described here represent a highly idealized model with representative values assumed for many parameters, we are encouraged by this agreement.

\begin{figure}[hbt!]
\epsscale{0.9}
\vspace{0.5in}
\plotone{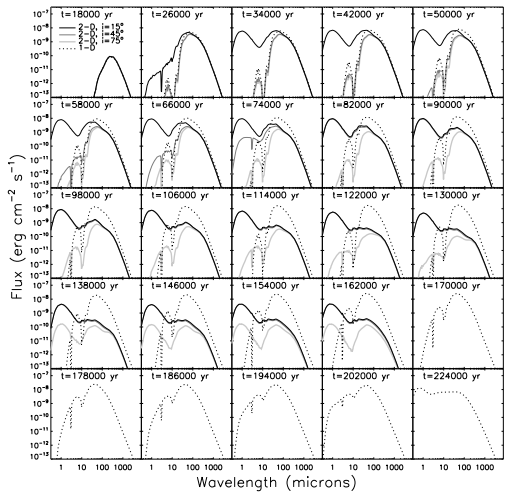}
\caption{\label{fig_mod4_seds}Same as Figure \ref{fig_mod3_seds}, except now for model 4: SEDs at select ages through the collapse of the 1 \msun\ core.  In each panel, the lines show the model 4 (scattering, disk, rotationally flattened envelope, and mass-loss and outflow cavities included) SEDs at various inclinations, with the line weight key given in the first panel.  The dotted line shows, for comparison, the SED at that time for model 1.  There are no model 4 SEDs at late times since collapse ends at 165,000 yr in the model 4 1 \msun\ initial mass core.}
\end{figure}

Figure \ref{fig_mod4_seds} compares the model 4 SEDs at various inclinations for the 1 \msun\ initial mass core to the model 1 SEDs presented in \S \ref{mod1}.  There are no model 4 SEDs at late times since collapse ends at 165,000 yr rather than 224,000 yr, as dicussed above.  The most striking change compared to models $1-3$ is the increased inclination dependence.  Once $\theta_c$ exceeds the inclination of a given line-of-sight the emission from the protostar and disk are directly observed, along with the long-wavelength emission from the envelope.  A line-of-sight's transition from passing through the envelope to passing through the outflow cavity is not immediate; there is a short transition as $\theta_c$ approaches that line-of-sight's inclination where it passes through both the cavity and the envelope, a result of the stream-line, funnel-shaped outflow cavity.  An example of this is seen in the 58,000 and 66,000 yr panels; the $i=45$\degree\ line-of-sight clearly shows an increase relative to the $i=75$\degree\ line-of-sight due to geometry even though $\theta_c$ does not reach 45\degree\ until 71,000 yr.  Finally, we also note that there is less long-wavelength emission at a given time compared to model 1 and that this discrepancy increases with time.  Since emission at these wavelengths directly traces total mass, this discrepancy is due to the faster decrease in $M_{env}$ induced by the entrainment of envelope material in the outflow.

\begin{figure}[hbt!]
\epsscale{0.6}
\vspace{0.35in}
\plotone{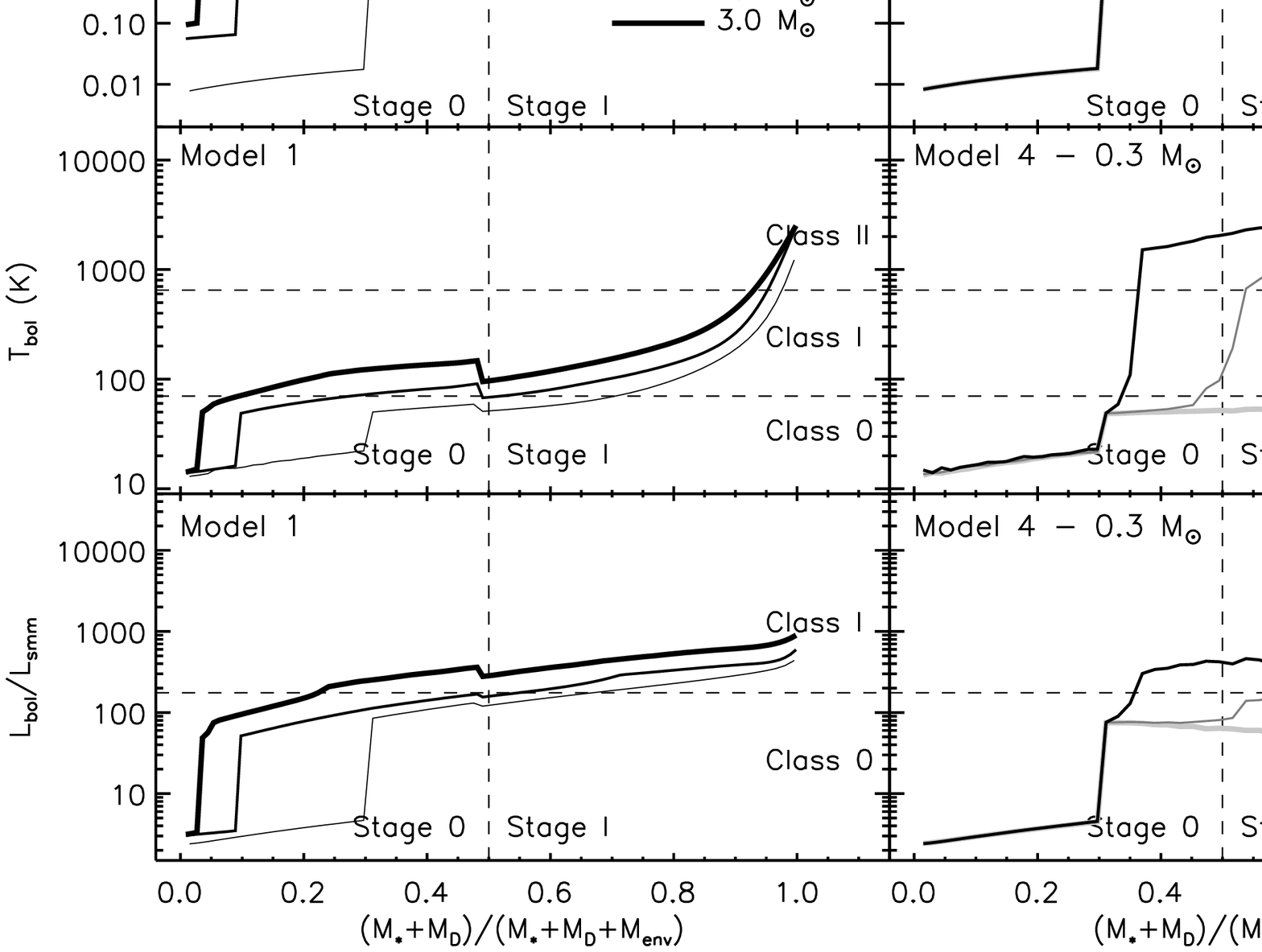}
\caption{\label{fig_mod4_evols1}Same as Figure \ref{fig_mod3_evols1}, except now for model 4: observational signatures \lbol\ (top panels), \tbol\ (middle panels) and \lbolsmm\ (bottom panels) versus $M_{int}/M_{tot}$, the ratio of internal (protostar+disk) to total (protostar+disk+envelope) mass.  The left panels show the model 1 results while the right panels show the model 4 results.  The different lines show the results for different inclinations; the line weight key is given in the upper right panel.  Only the 0.3 \msun\ model 4 core is shown to avoid creating an overly complicated figure.  The class boundaries in \tbol\ are taken from Chen et al.~(1995) while the class divisions in \lbolsmm\ are taken from YE05.  The discontinuities in both \tbol\ and \lbolsmm\ for model 1 are artifacts introduced by the switch from the Shu (1977) density profile to a power-law density profile.}
\end{figure}

\begin{figure}[t]
\epsscale{0.6}
\vspace{0.35in}
\plotone{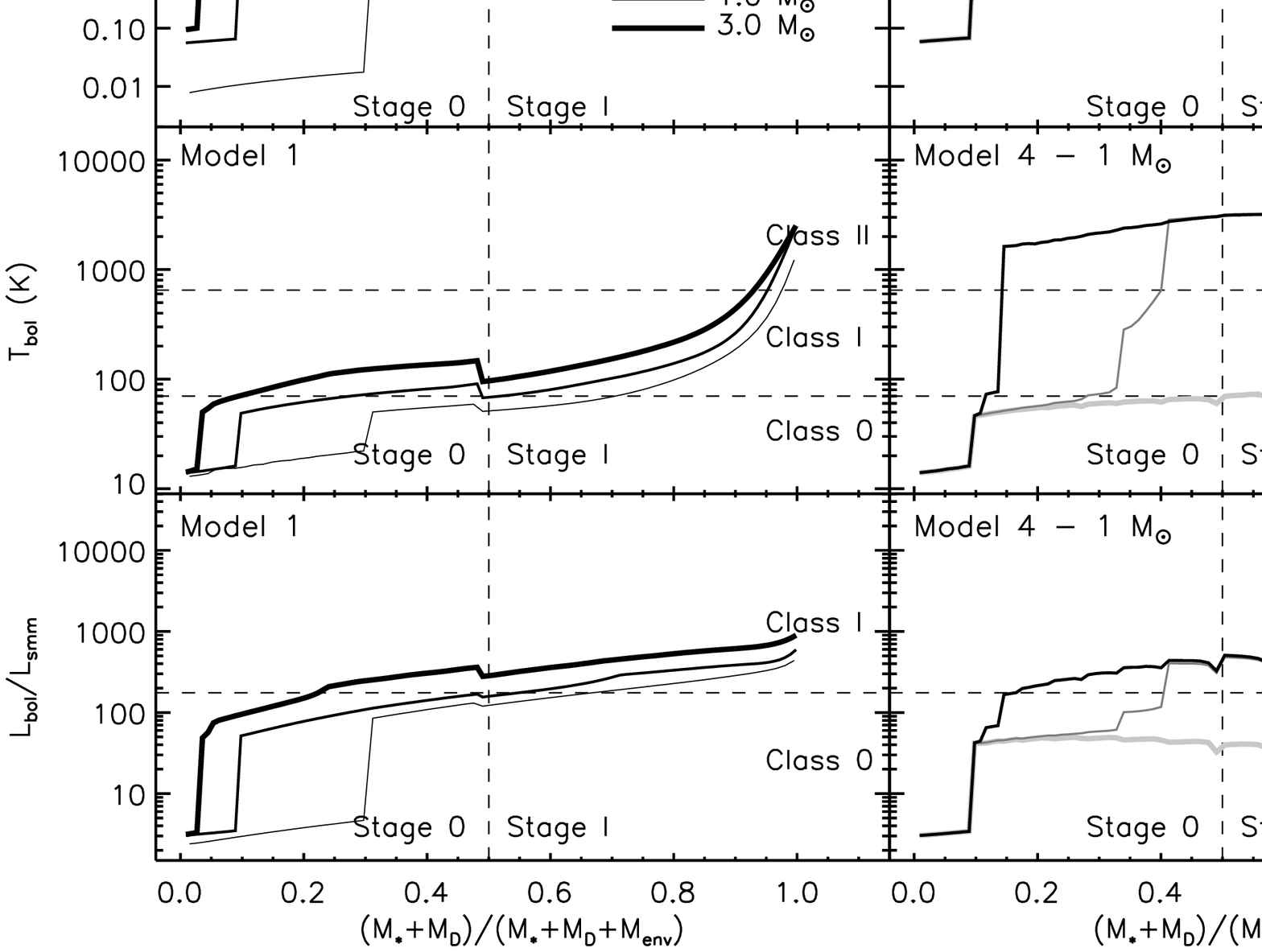}
\caption{\label{fig_mod4_evols2}Same as Figure \ref{fig_mod4_evols1}, except now showing the 1 \msun\ rather than the 0.3 \msun\ model 4 core: observational signatures \lbol\ (top panels), \tbol\ (middle panels) and \lbolsmm\ (bottom panels) versus $M_{int}/M_{tot}$, the ratio of internal (protostar+disk) to total (protostar+disk+envelope) mass.  The left panels show the model 1 results while the right panels show the model 4 results.  The different lines show the results for different inclinations; the line weight key is given in the upper right panel.  The class boundaries in \tbol\ are taken from Chen et al.~(1995) while the class divisions in \lbolsmm\ are taken from YE05.  The discontinuities in both \tbol\ and \lbolsmm\ for model 1 are artifacts introduced by the switch from the Shu (1977) density profile to a power-law density profile.}
\end{figure}

\begin{figure}[hbt!]
\epsscale{0.6}
\vspace{0.35in}
\plotone{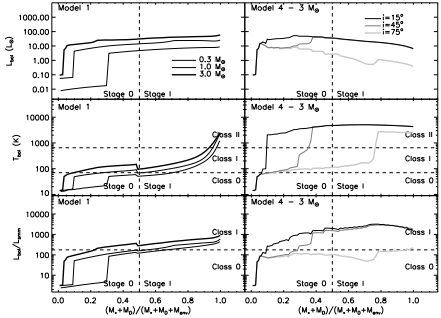}
\caption{\label{fig_mod4_evols3}Same as Figure \ref{fig_mod4_evols2}, except now showing the 3 \msun\ rather than the 1 \msun\ model 4 core: observational signatures \lbol\ (top panels), \tbol\ (middle panels) and \lbolsmm\ (bottom panels) versus $M_{int}/M_{tot}$, the ratio of internal (protostar+disk) to total (protostar+disk+envelope) mass.  The left panels show the model 1 results while the right panels show the model 4 results.  The different lines show the results for different inclinations; the line weight key is given in the upper right panel.  The class boundaries in \tbol\ are taken from Chen et al.~(1995) while the class divisions in \lbolsmm\ are taken from YE05.  The discontinuities in both \tbol\ and \lbolsmm\ for model 1 are artifacts introduced by the switch from the Shu (1977) density profile to a power-law density profile.}
\end{figure}

Figures \ref{fig_mod4_evols1}, \ref{fig_mod4_evols2}, and \ref{fig_mod4_evols3} show the observational signatures \lbol, \tbol, and \lbolsmm\ plotted against the ratio of $M_{int}/M_{tot}$ for the 0.3, 1, and 3 \msun\ initial mass cores, respectively.  Unlike for previous models we do not combine the three masses on one plot since the increased inclination dependence of model 4 compared to models $1-3$ would create an overly complicated, difficult-to-read figure.  The results discussed in relation to Figure \ref{fig_mod4_seds} are readily apparent: a given line-of-sight features low values of \tbol\ and \lbolsmm\ until $\theta_c$ approaches the inclination of that line-of-sight, and after a small transition region where both quantities increase as the line-of-sight passes through more of the cavity and less of the envelope, they increase to high values characteristic of those expected for viewing a protostar+disk directly through the outflow cavity.  The calculated \lbol\ also shows an inclination dependence and is generally lower than in previous models due to the lower protostellar masses and lower mass accretion rates as a consequence of including mass-loss.

Unlike with previous models, it no longer makes sense to quote an inclination-weighted average value of $M_{int}/M_{tot}$ when each initial mass core crosses the Class 0/I boundary in either \tbol\ or \lbolsmm.  Indeed, the 0.3 \msun\ initial mass core crosses the \tbol\ Class 0/I boundary at values of $M_{int}/M_{tot}$ ranging from 0.31 to 0.91 depending on inclination, and it crosses the \lbolsmm\ Class 0/I boundary at values of $M_{int}/M_{tot}$ ranging from 0.33 to never,\footnote{Even when the envelope has fully dissipated, the disk keeps the nearly edge-on SEDs from crossing the Class 0/I boundary in \lbolsmm.  The reason why this did not occur for models 2 and 3, which also included a disk in the radiative transfer, is because of the lower \lbol\ in model 4 due to lower protostellar masses and lower accretion rates.  \lsmm\ stays about the same, but \lbol\ decreases, lowering \lbolsmm\ in model 4 compared to models 2 and 3.} depending on inclination.  Similar results are found for the 1 and 3 \msun\ initial mass cores.  The entire concept of a connection between physical Stage and observational Class as measured by \tbol\ or \lbolsmm\ breaks down once the inclination dependence from outflow cavities are taken into account, since either quantity can vary by an order of magnitude or more depending on inclination.  These results are in general agreement with Crapsi et al.~(2008), who found that \tbol\ can vary by factors of $\sim 2-6$ depending on the exact model parameters.  However, as their models held $\theta_c$ fixed at 15\degree\ but adopted $i=25$\degree\ as their minimum inclination, their maximum variation in \tbol\ does not include a line-of-sight looking directly down the outflow cavity.  With such small cavities, they concluded that \tbol\ still provided a good measure of physical Stage for moderate inclinations ($25-70$\degree) with lines-of-sight that do not pass through either the cavity or the disk.  On the contrary, we find that neither \tbol\ nor \lbolsmm\ provides a good measure of physical Stage regardless of inclination.

\begin{figure}[hbt!]
\epsscale{0.90}
\plotone{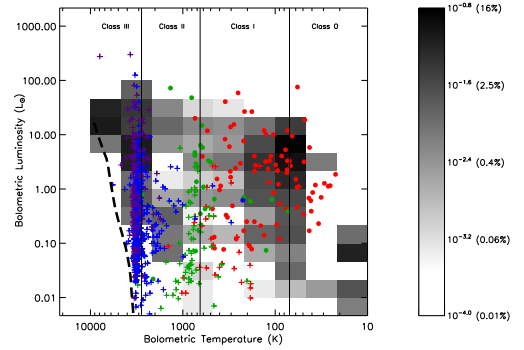}
\caption{\label{fig_mod4_blt}Same as Figure \ref{fig_mod3_blt}, except now for model 4.  The grayscale pixels indicate the fraction of total time the model spends in each \lbol\ $-$ \tbol\ bin, calculated from Equation \ref{eq_bins}.  The grayscale is displayed in a logarithmic stretch with the scaling chosen to emphasize the full extent of the models in \lbol\ $-$ \tbol\ space.  The mapping between grayscale and fraction of total time is indicated in the legend.  The class boundaries in \tbol\ are taken from Chen et al.~(1995).  The thick dashed line shows the ZAMS (D'Antona \& Mazzitelli 1994) from 0.1 to 2.0 \msun.  The colored symbols show the Young Stellar Objects from Evans et al.~(2009) in this diagram; the colors and symbols hold the same meaning as in Figure \ref{fig_mod1_blt}.}
\end{figure}

\begin{figure}[hbt!]
\epsscale{1.0}
\plotone{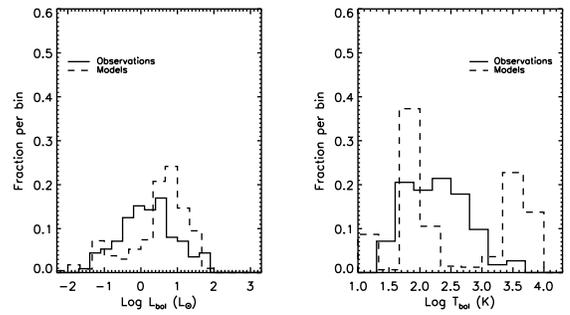}
\caption{\label{fig_mod4_hist}Same as Figure \ref{fig_mod3_hist}, except now for model 4: histograms showing the fraction of total sources (observations; solid lines) and fraction of total time spent (models; dashed lines; calculated from Equation \ref{eq_bins}) at various \lbol\ (left) and \tbol\ (right).  The binsize is 1/3 dex in both quantities.  For the observations, only the 112 embedded sources (plotted as filled circles on the BLT diagrams) are included.}
\end{figure}

Figure \ref{fig_mod4_blt} shows a BLT diagram for model 4.  Unlike with previous models, the full extent of the embedded sources in \lbol\ $-$ \tbol\ space is reproduced by model 4.  However, Figure \ref{fig_mod4_hist}, which plots model 4 \lbol\ and \tbol\ histograms, shows that while the distribution of time spent at various luminosities is wider in model 4 than in models $1-3$ and clearly gives a better fit to the observed distribution (a K-S test gives a 22\% probability that the observed and model \lbol\ histograms represent the same underlying distribution, compared to $< 0.1$\% for models $1-3$), the model still overpredicts the time spent at $\sim 2-20$ \lsun\ and underpredicts the time spent at $\sim 0.1-2$ \lsun.  Figure \ref{fig_mod4_blt} also shows that the model spends a relatively large fraction of time at high \tbol\ ($\ga 1000$ K) compared to the fraction of embedded sources at such values, a consequence of viewing direct protostar+disk emission though outflow cavities for many model lines-of-sight.  This is also evident in both Figure \ref{fig_mod4_hist} (a K-S test gives a 16\% probability that the observed and model \tbol\ histograms represent the same underlying distribution) and column 7 of Table \ref{tab_fractionperbin}, which lists the fraction of total time model 4 spends in various \lbol\ $-$ \tbol\ bins.  The model spends 40.1\% of the time at \tbol\ $\geq 1000$ K whereas only 4.5\% of the embedded sources are found at such high \tbol.  The model spends most of the rest of the time at low \tbol\ ($\la 100$ K) whereas the embedded observations are relatively evenly distributed (in a logarithmic binning) between $\sim 20-500$ K.

In summary, the main conclusions of models $1-3$ that the model overpredicts both the time spent at high \lbol\ ($\ga 1-2$ \lsun) and the time spent at low \tbol\ ($\la 100-200$ K) remain unchanged.  However, including the effects of mass-loss and outflow cavities has reduced the severity of the luminosity problem and also introduced a significant population of model 4 SEDs with higher \tbol\ than found for embedded sources.

Before moving on, we briefly return to the assumed value of $f$, the efficiency with which the jet/wind transfers its momentum to the ambient medium.  We assumed representative values based on either observed or theoretical ranges for all other relevant parameters, but we maximized $f$ by setting it equal to 1.  We made this choice to maximize the effects of mass-loss and outflow cavities, since $f=1$ maximizes the amount of mass entrained in the outflow and thus the increase in $\theta_c$ with time.  Even with their effects maximized, mass-loss and outflow-cavities still feature the same basic luminosity problem as the other models, albeit to a lesser degree.  We consider this to be a strong test of the necessity of invoking episodic accretion to explain the observed distribution of embedded sources in \lbol\ $-$ \tbol\ space.  However, with $f=1$, $\theta_c$ reaches 90\degree\ before collapse ends (defined as $t_{collapse} = M_{env}^{initial} / \dot{M}_{env}^{initial}$) and thus terminates the embedded phase earlier than when no outflow cavities are present, and the large cavities prevent both \tbol\ and \lbolsmm\ from being useful indicators of physical Stage.  What if we had assumed a more common value in the range of $f=0.1-0.25$ (e.g., \andre\ et al.~1999)?

With $f=0.1$, collapse ends before $\theta_c$ reaches 90\degree\ in all three initial mass cores.  However, these angles still reach 60, 65, and 75\degree\ at the end of the collapse of the 0.3, 1, and 3 \msun\ initial mass cores, respectively.  Thus, we still disagree with the conclusion of Crapsi et al.~(2008) that \tbol\ provides a good measure of evolutionary status for moderate inclinations ($25-70$\degree) with lines-of-sight that do not pass through either the cavity or the disk, which they reached by assuming a constant $\theta_c = 15$\degree.  Furthermore, the measured values of $f_{SFE}$ with $f=0.1$ are 0.72, 0.64, and 0.64, respectively, much higher than those measured with $f=1$ and found above to be in general agreement with observational estimates of this quantity.  Finally, $f=0.1$ means that the protostellar mass will grow more quickly and the envelope accretion rate will decrease more slowly, both of which will increase model luminosities that are either dominated by or include significant components from accretion luminosity (see \S \ref{model_luminosities}).  We thus conclude that choosing more common values of $0.1-0.25$ for the entrainment efficiency will not change any of our basic results, and will lessen the degree to which the luminosity problem is improved by model 4.

\subsection{Model 5: Episodic Accretion}\label{mod5}

In models $1-4$ we have modified the YE05 model to include isotropic scattering off dust grains, a two-dimensional disk in the radiative transfer, rotationally flattened envelope density profiles following the TSC84 solution for the collapse of slowly rotating cores, and mass-loss and outflow cavities.  All of these have had impacts on the model predictions, especially including the opacity from scattering and mass-loss and outflow cavities.  However, even with all the above effects considered, the models still exhibit the fundamental luminosity problem described in \S \ref{intro}: the models overpredict the fraction of total time spent at high luminosities ($\ga 1-2$ \lsun) compared to observations of embedded protostars (although the effects of mass-loss and outflow cavities reduce the severity of the problem).  Given this, along with the discussion in \S \ref{intro} regarding observational evidence for non-steady mass accretion and theoretical predictions of episodic mass accretion, we incorporate episodic mass accretion into our evolutionary model.  As with \S \ref{mod4}, we incorporate a simple, idealized process of episodic accretion that, while physically motivated, is designed to test its general effects on the evolutionary signatures of embedded protostars rather than fully capture a complete, physically self-consistent model.

The physical basis for our treatment of episodic accretion are the models published by Vorobyov \& Basu (2005b, 2006), who use MHD simulations to follow the collapse of rotating cores.  In their simulations, material piles up in a circumstellar disk until the disk becomes gravitationally unstable and develops spiral structure and dense clumps, which are then driven onto the protostar in short-lived accretion bursts generated through the gravitational torques associated with the spiral arms.  They found this burst phenomenon to be a robust result under a variety of initial conditions (Vorobyov \& Basu 2006).  Other authors have also shown that gravitational instabilities in the disk can produce episodic mass accretion onto the protostar.  For example, Boss (2002) noted that their models of gravitationally unstable disks (used primarily to investigate giant planet formation) feature protostellar mass accretion rates that vary with time between about $10^{-7}$ to $10^{-3}$ \msun\ yr$^{-1}$.

To incorporate episodic accretion into our model, we allow the envelope to evolve as before: material accretes from the envelope directly onto the protostar and disk (but mostly the disk as it grows in size), with the envelope density profile evolving following the TSC84 solution for the collapse of slowly rotating cores.  This leaves the 1st, 2nd, and 4th luminosity sources as described in \S \ref{model_luminosities} unchanged.  However, instead of material accreting through the disk and onto the protostar at the rate $\dot{M}_{DtoP} = \eta_D \dot{M}_{env}$, all material accreted onto the disk is stored in the disk ($\dot{M}_{DtoP}$ is set to zero).  This causes the 3rd luminosity source, that arising from accretion from the disk onto the protostar, to vanish, and the 5th luminosity source, which depends on the total mass accretion rate onto the protostar, to be significantly reduced.  Since the luminosity arising from accretion from the disk onto the protostar is the dominant source of luminosity throughout most of the model evolution (due to the deep potential well of the protostar), this will significantly reduce the total model luminosity.

The disk mass can't continue to grow indefinitely; eventually gravitational instabilities will set in once its mass grows to a significant fraction of the protostellar mass.  In a study of disks around Class II sources in Ophiuchus and Taurus, Andrews \& Williams (2007) found a $M_d/M_*$ distribution ranging from $\sim 10^{-3} - 0.2$.  If we assume the upper end of this distribution represents the youngest disks that have not yet begun to disperse, this should be a good estimate of the maximum possible $M_d/M_*$ as any disk higher than this would have become gravitationally unstable.  This is in good agreement with theoretical predictions of when gravitational instabilities develop ($M_d/M_* \sim 0.2$; e.g., Shu et al.~1990), and also with Vorobyov \& Basu (2006), who showed that the disk mass always remains significantly less than the protostellar mass in their simulations.  More recently, Vorobyov (2009a) argued that the maximum $M_d/M_*$ is closer to $0.25-0.4$; such a higher value would increase the time between bursts but also the duration of each burst, and would not significantly change our results.

Thus, once $M_d/M_*$ reaches 0.2, the model enters a burst mode and $\dot{M}_{DtoP}$ increases to $1 \times 10^{-4}$ \msun\ yr$^{-1}$.  In their simulations, Vorobyov \& Basu (2005b, 2006) found that $\dot{M}_{DtoP}$ increases to about ($1-5$) $\times 10^{-4}$ \msun\ yr$^{-1}$, even reaching $\sim 10^{-3}$ \msun\ yr$^{-1}$ in the most extreme cases; here we assume a constant value of $1 \times 10^{-4}$ \msun\ yr$^{-1}$ for every burst for simplicity.  In the bursts we increase the time resolution from $\Delta t$ = 1000, 2000, or 6000 yr for the 0.3, 1, and 3 \msun\ initial mass cores, respectively, to $\Delta t$ = 100 yr.  The high $\dot{M}_{DtoP}$ causes both the 3rd and 5th luminosity sources as described in \S \ref{model_luminosities} to increase to very high values, dominating the total luminosity during bursts (see below).  The protostellar mass grows rapidly during a burst, and the disk mass decreases accordingly (during a burst the material is accreting out of the disk about two orders of magnitude faster than it is accreting onto the disk from the envelope).  For simplicity the burst is assumed to continue until all of the disk mass accretes onto the protostar, at which point $\dot{M}_{DtoP}$ drops back to zero, the time resolution decreases back to its original value, and the cycle begins anew.  We assume the dust temperature has reached equilibrium by the timesteps immediately following the onset and termination of accretion bursts\footnote{Once the change in luminosity reaches the dust, it will respond very quickly to the change, reaching its new equilibrium within a few seconds (e.g., Draine \& Anderson 1985; Lee et al.~2007).  The random-walk time through the envelope is the limiting factor, and a conservative upper limit calculated assuming photons remain at their initial wavelength instead of being reprocessed to longer wavelengths (where the optical depth is lower) puts this timescale at $\sim 100$ yr.  In reality it is much less once reprocessing to longer wavelengths is considered.}.

\begin{figure}[hbt!]
\epsscale{0.6}
\vspace{0.5in}
\plotone{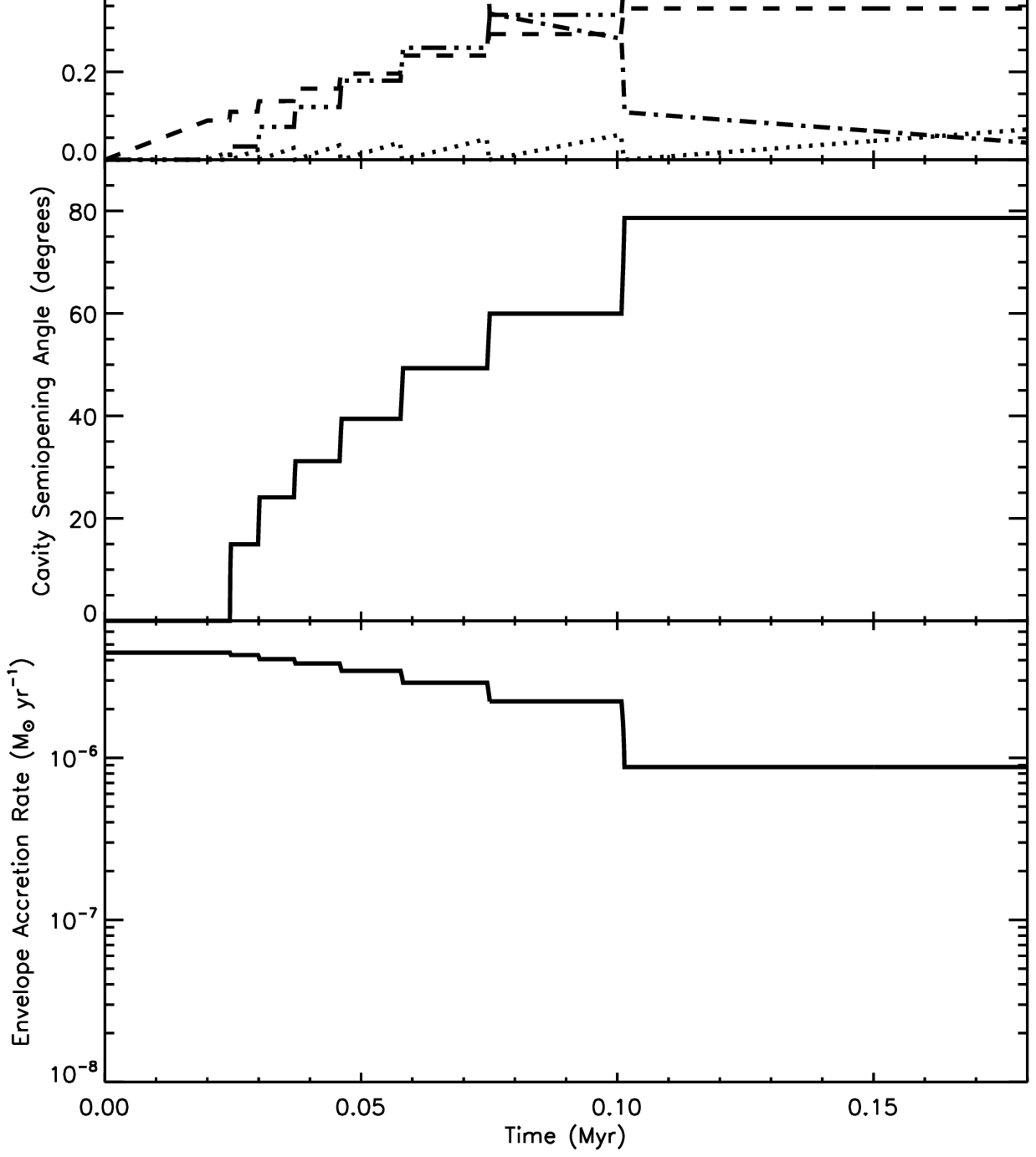}
\caption{\label{fig_mod5_setup}Same as Figure \ref{fig_mod4_setup}, except now for model 5: evolution of the masses of various model components (top panel), outflow cavity semi-opening angle (middle panel), and envelope mass accretion rate (bottom panel) versus time for the 1 \msun\ initial mass model 5 core, which includes episodic accretion as described in \S \ref{mod5}.}
\end{figure}

Figure \ref{fig_mod5_setup} shows the effects of including episodic accretion as described above for the 1 \msun\ initial mass core.  Similar to Figure \ref{fig_mod4_setup} for model 4, plotted are the masses of the various model components (protostar, disk, envelope, and outflow), the cavity semi-opening angle, and $\dot{M}_{env}$ versus time.  Except for the first 0.02 Myr when the disk has not yet formed, $M_*$ shows a ``staircase'' function, essentially increasing only during the bursts (except for the very small amount of accretion directly from the envelope onto the star).  Each increase in $M_*$ is accompanied by a corresponding decrease in $M_d$.  Each burst also features an increase in the outflow mass and decrease in $M_{env}$ since mass is ejected (and thus envelope mass is entrained) when material accretes from the protostar to the disk.  These increases in the outflow mass are accounted for by increases in the outflow cavity semi-opening angle and thus cause decreases in $\dot{M}_{env}$ following Equation \ref{eq_mdotnew}.  The collapse ends at 180,000 yr; a burst begins at this time and by the next timestep, 100 yr later, the outflow cavity semi-opening angle has reached 90\degree, removing the remaining envelope material and ending collapse.

The 0.3, 1, and 3 \msun\ initial mass cores feature 3, 8, and 11 bursts, respectively.  Vorobyov \& Basu (2005b) found $15-30$ bursts in their initial simulation, but showed in Vorobyov \& Basu (2006) that the exact number depends on assumed values of both $\Omega_0$ and the magnetic field.  It will also depend on the exact criterion assumed for gravitational instability, which we have held fixed at $M_d/M_* = 0.2$ for simplicity.  All three initial mass cores spend $\sim 1.5-2$\% of their total collapse times in burst phases and have $f_{SFE}$ of 0.53, 0.35, and 0.33, respectively.  As expected, these are similar to model 4 since including episodic accretion changes when material accretes onto the protostar but does not generally affect the total amount of material accreted.

\begin{figure}[hbt!]
\epsscale{0.9}
\plotone{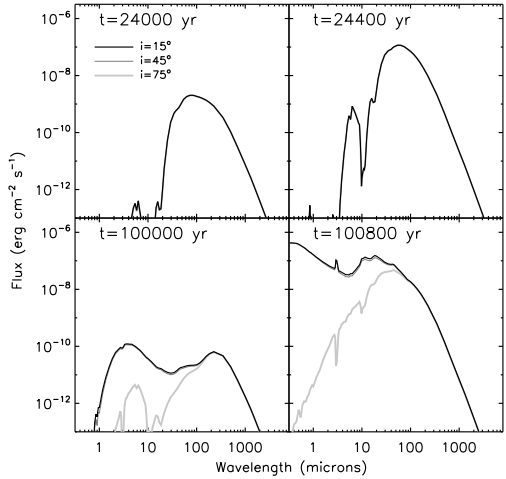}
\caption{\label{fig_mod5_seds}Select SEDs in the collapse of the model 5 1\msun\ core.  The top panels show the SEDs just before ($t=24000$ yr; right) and during ($t=24400$ yr; left) an accretion burst that begins at $t=24400$ yr.  The bottom panels show the SEDs just before ($t=100000$ yr; right) and during ($t=100800$ yr; left) an accretion burst that begins at $t=100800$ yr.  Three different inclinations are shown: nearly pole-on ($i=15$\degree), moderate ($i=45$\degree), and nearly edge-on ($i=75$\degree), with the line weight key given in the first panel.  In the top two panels all three inclinations feature nearly equivalent SEDs since the departure from spherical symmetry is very small at early times.  In the bottom two panels the nearly pole-on and moderate inclinations feature nearly equivalent SEDs because they are both viewing direct protostar+disk emission through the outflow cavity.}
\end{figure}

Showing SEDs at select ages through the collapse of the 1 \msun\ initial mass core is less meaningful here than for models $1-4$ since the evolution from starless core to revealed protostar+disk is no longer smooth but changes abruptly in the bursts.  We instead show, in Figure \ref{fig_mod5_seds}, the SEDs just before and during accretion bursts that begin at $t=24400$ yr and $t=100800$ yr.  While the core itself is at a much different evolutionary state at the two times, the effects of an accretion burst are similar: the flux increases due to the increase in luminosity, and the overall shape of the SED shifts to shorter wavelengths due to the increase in emission from both the protostar itself and the warm dust in the envelope heated by the protostar.

\begin{figure}[hbt!]
\epsscale{0.6}
\vspace{0.35in}
\plotone{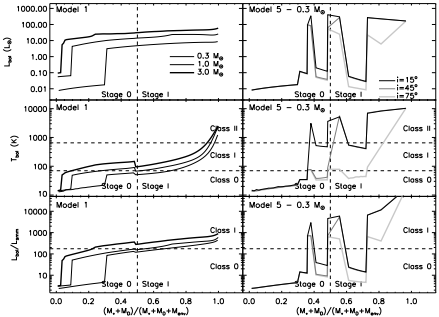}
\caption{\label{fig_mod5_evols1}Same as Figure \ref{fig_mod4_evols1}, except now for model 5: observational signatures \lbol\ (top panels), \tbol\ (middle panels) and \lbolsmm\ (bottom panels) versus $M_{int}/M_{tot}$, the ratio of internal (protostar+disk) to total (protostar+disk+envelope) mass.  The left panels show the model 1 results while the right panels show the model 5 results.  The different lines show the results for different inclinations; the line weight key is given in the upper right panel.  Only the 0.3 \msun\ model 5 core is shown to avoid creating an overly complicated figure.  The class boundaries in \tbol\ are taken from Chen et al.~(1995) while the class divisions in \lbolsmm\ are taken from YE05.  The discontinuities in both \tbol\ and \lbolsmm\ for model 1 are artifacts introduced by the switch from the Shu (1977) density profile to a power-law density profile.}
\end{figure}

\begin{figure}[hbt!]
\epsscale{0.6}
\vspace{0.35in}
\plotone{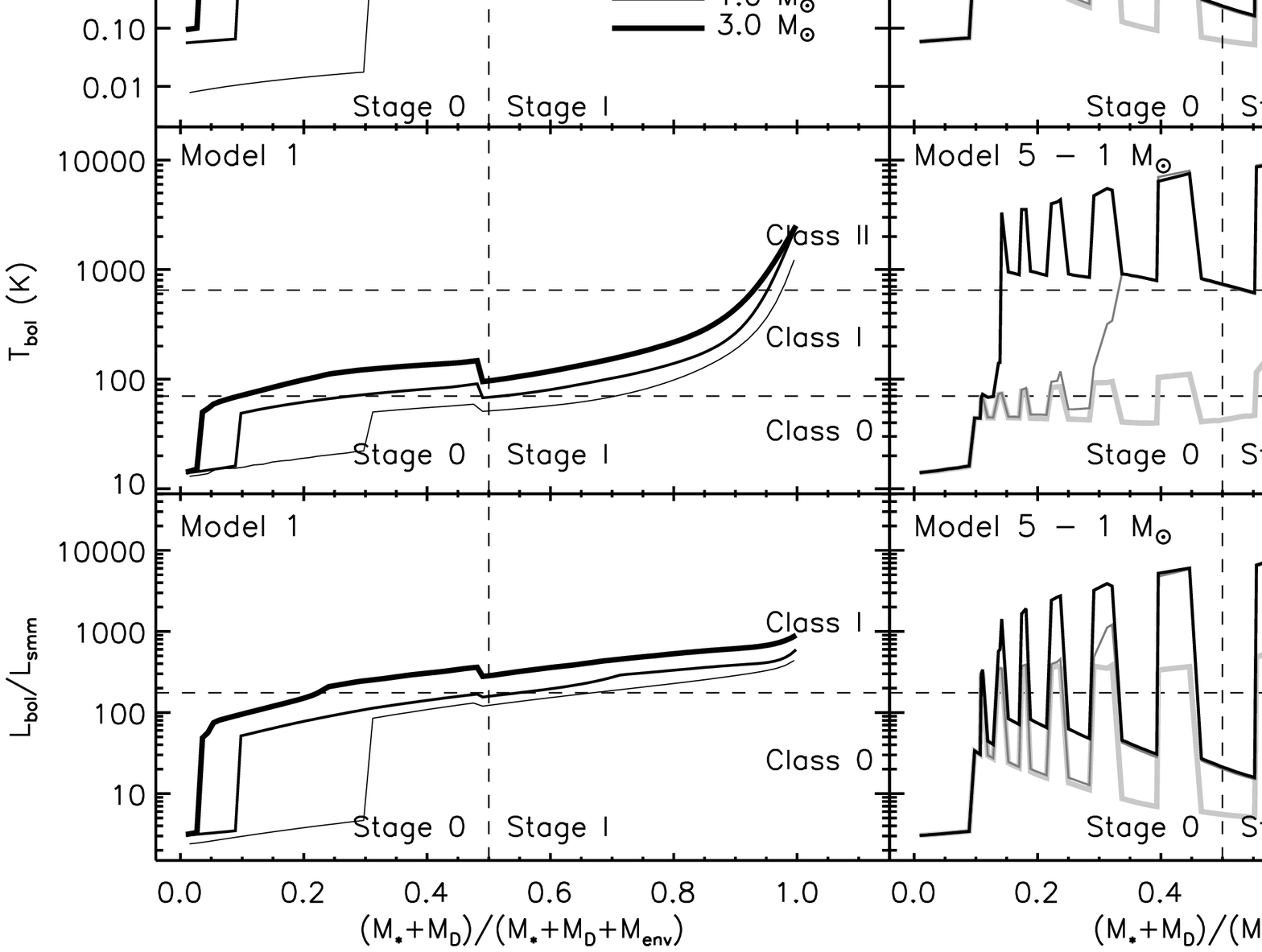}
\caption{\label{fig_mod5_evols2}Same as Figure \ref{fig_mod5_evols2}, except now showing the 1 \msun\ rather than the 0.3 \msun\ model 5 core: observational signatures \lbol\ (top panels), \tbol\ (middle panels) and \lbolsmm\ (bottom panels) versus $M_{int}/M_{tot}$, the ratio of internal (protostar+disk) to total (protostar+disk+envelope) mass.  The left panels show the model 1 results while the right panels show the model 5 results.  The different lines show the results for different inclinations; the line weight key is given in the upper right panel.  The class boundaries in \tbol\ are taken from Chen et al.~(1995) while the class divisions in \lbolsmm\ are taken from YE05.  The discontinuities in both \tbol\ and \lbolsmm\ for model 1 are artifacts introduced by the switch from the Shu (1977) density profile to a power-law density profile.}
\end{figure}

\begin{figure}[hbt!]
\epsscale{0.6}
\vspace{0.35in}
\plotone{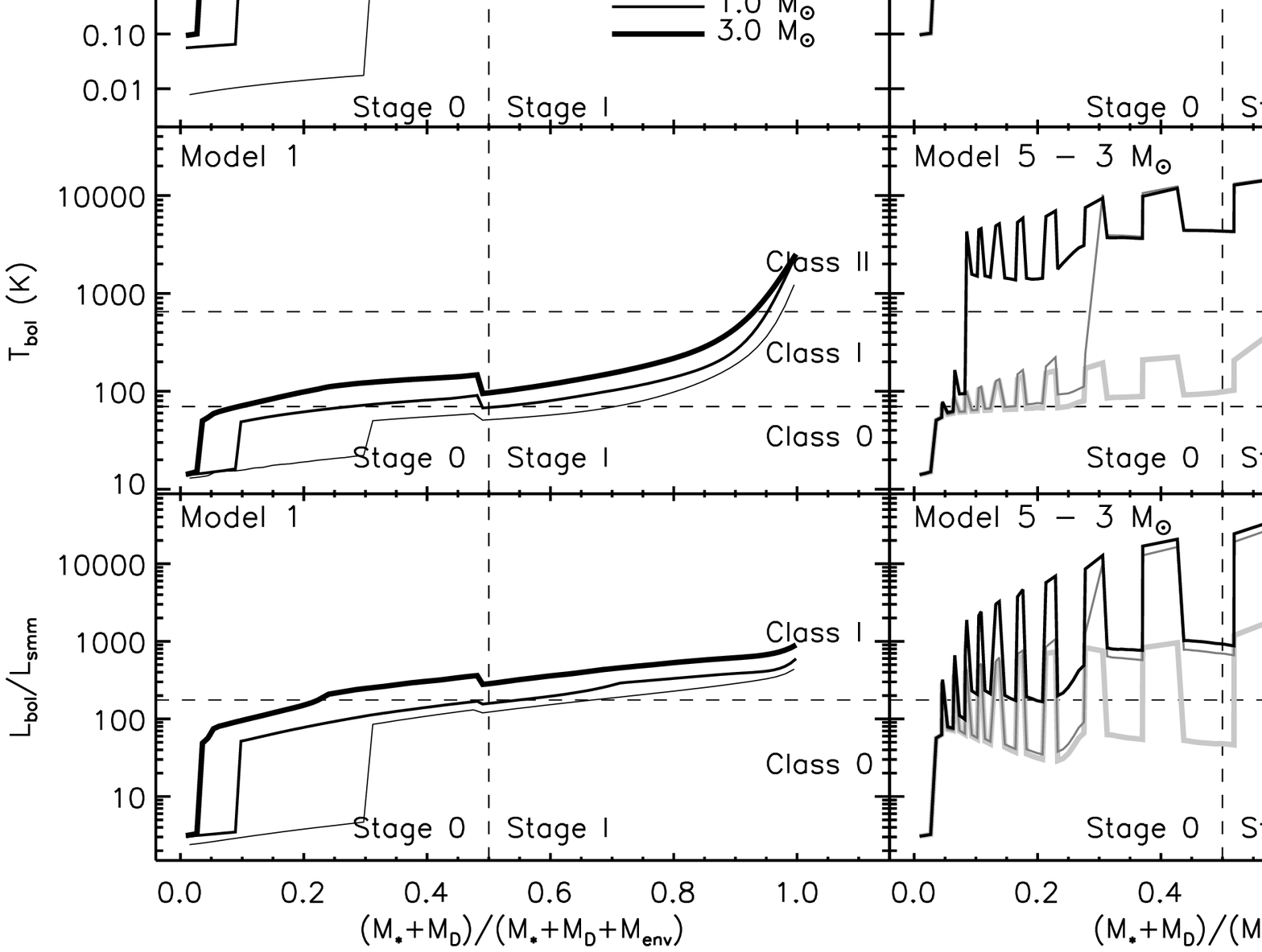}
\caption{\label{fig_mod5_evols3}Same as Figure \ref{fig_mod5_evols2}, except now showing the 3 \msun\ rather than the 1 \msun\ model 5 core: observational signatures \lbol\ (top panels), \tbol\ (middle panels) and \lbolsmm\ (bottom panels) versus $M_{int}/M_{tot}$, the ratio of internal (protostar+disk) to total (protostar+disk+envelope) mass.  The left panels show the model 1 results while the right panels show the model 5 results.  The different lines show the results for different inclinations; the line weight key is given in the upper right panel.  The class boundaries in \tbol\ are taken from Chen et al.~(1995) while the class divisions in \lbolsmm\ are taken from YE05.  The discontinuities in both \tbol\ and \lbolsmm\ for model 1 are artifacts introduced by the switch from the Shu (1977) density profile to a power-law density profile.}
\end{figure}

Figures \ref{fig_mod5_evols1}, \ref{fig_mod5_evols2}, and \ref{fig_mod5_evols3} show the observational signatures \lbol, \tbol, and \lbolsmm\ plotted against the ratio of $M_{int}/M_{tot}$ for the 0.3, 1, and 3 \msun\ initial mass cores, respectively.  As with model 4, we show each initial mass core in a separate figure to avoid creating a confusing, difficult-to-read figure.  Since the envelope mass quickly decreases in each burst due to the entrainment of material in the outflow (see above and Figure \ref{fig_mod5_setup}), $M_{tot}$ also quickly decreases in each burst and thus the ratio of $M_{int}/M_{tot}$ quickly increases.  As a result, the values on the x-axis in Figures \ref{fig_mod5_evols1} $-$ \ref{fig_mod5_evols3} no longer increase linearly with time; the bursts occupy a much lower fraction of total time than they do total $M_{int}/M_{tot}$.

Without accretion from the disk onto the protostar providing the dominant source of luminosity, most of the time is spent at low luminosity (often $\geq 1-2$ orders of magnitude lower compared to model 1) except during the bursts, when the luminosity increases to very high values ($\sim 100-1000$ \lsun).  The general conclusion from model 4 that the large inclination dependence introduced by the outflow cavities prevents both \tbol\ and \lbolsmm\ from providing good measures of evolutionary status remains unchanged.  For example, excluding time spent in bursts, the 1 \msun\ initial mass core crosses the \tbol\ Class 0/I boundary at values of $M_{int}/M_{tot}$ ranging from $0.11 - 0.90$ depending on inclination.  Furthermore, again excluding time spent in bursts, this core spends essentially all of its time as Class 0 by \lbolsmm, crossing the Class 0/I boundary at values of $M_{int}/M_{tot}$ ranging from $0.84 -$ never depending on inclination.  Both quantities are similarly unreliable at measuring the evolutionary status of the 0.3 and 3 \msun\ initial mass core.

However, episodic accretion adds another layer of unreliability to these evolutionary indicators.  Both quantities clearly show large increases during bursts and subsequent decreases after each burst ends.  Depending on the combination of inclination and age, \tbol\ and \lbolsmm\ calculated for a given line-of-sight can cross back and forth across the Class 0/I boundary in both quantities several times throughout the collapse of the core.  If this model represents physical reality, which we evaluate below, neither of the commonly used evolutionary indicators actually reliably trace physical Stage.

\begin{figure}[hbt!]
\epsscale{0.90}
\plotone{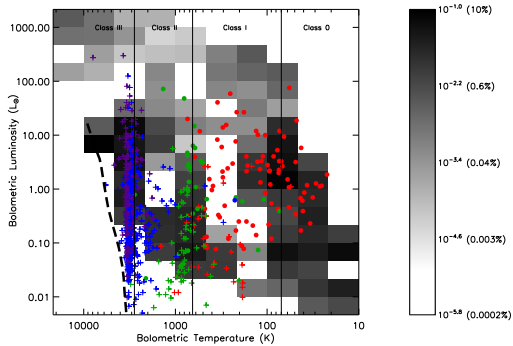}
\caption{\label{fig_mod5_blt}Same as Figure \ref{fig_mod4_blt}, except now for model 5.  The grayscale pixels indicate the fraction of total time the model spends in each \lbol\ $-$ \tbol\ bin, calculated from Equation \ref{eq_bins}.  The grayscale is displayed in a logarithmic stretch with the scaling chosen to emphasize the full extent of the models in \lbol\ $-$ \tbol\ space.  The mapping between grayscale and fraction of total time is indicated in the legend.  The class boundaries in \tbol\ are taken from Chen et al.~(1995).  The thick dashed line shows the ZAMS (D'Antona \& Mazzitelli 1994) from 0.1 to 2.0 \msun.  The colored symbols show the Young Stellar Objects from Evans et al.~(2009) in this diagram; the colors and symbols hold the same meaning as in Figure \ref{fig_mod1_blt}.}
\end{figure}

Figure \ref{fig_mod5_blt} shows a BLT diagram for model 5.  The full extent of the embedded sources in \lbol\ $-$ \tbol\ space is again reproduced, as for model 4.  The model reaches higher luminosities than models $1-4$ as a result of the accretion bursts.  At first glance these high luminosities appear inconsistent with observations, although it is difficult to evaluate from this figure alone given the logarithmic grayscale.  The model only spends $\sim$1\% of its total time at \lbol\ $\geq$ 100 \lsun;\footnote{Even though each core spends about 2\% of the total collapse time in burst, and the total model luminosity in every burst is $>$ 100 \lsun, the higher inclination lines-of-sight that are weighted more heavily due to the increased solid angle at higher inclinations [see \S \ref{mod1}] often have \lbol\ $<$ 100 \lsun\ even in accretion bursts} for comparison, none of the observed sources have \lbol\ $\geq$ 100 \lsun.  Given that the observed sample only consists of 112 sources, $\sim$1\% only corresponds to about one source, and thus it is not inconsistent, given the small-number statistics, to find no sources at such high luminosities.

\begin{figure}[hbt!]
\epsscale{1.0}
\plotone{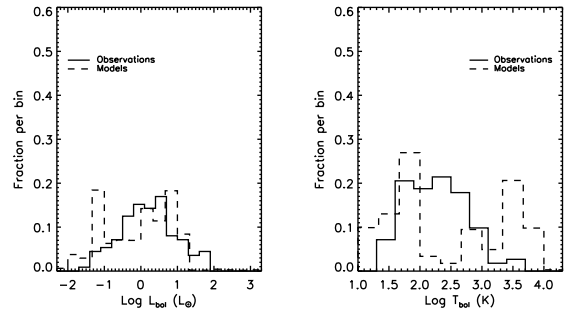}
\caption{\label{fig_mod5_hist}Same as Figure \ref{fig_mod4_hist}, except now for model 5: histograms showing the fraction of total sources (observations; solid lines) and fraction of total time spent (models; dashed lines; calculated from Equation \ref{eq_bins}) at various \lbol\ (left) and \tbol\ (right).  The binsize is 1/3 dex in both quantities.  For the observations, only the 112 embedded sources (plotted as filled circles on the BLT diagrams) are included.}
\end{figure}

Figure \ref{fig_mod5_hist} plots model 5 \lbol\ and \tbol\ histograms, and column 8 of Table \ref{tab_fractionperbin} lists the fraction of total time model 5 spends in various \lbol\ $-$ \tbol\ bins.  A K-S test shows that there is a 61\% probability that the observed and model \lbol\ histograms represent the same underlying distribution, by far the highest of models $1-5$.  Inspection of Figure \ref{fig_mod5_hist} shows that most of the remaining discrepancy between the models and observations is in the form of a sort of ``reverse luminosity problem'' in that the models spend an excess of time at \lbol\ $\sim$ 0.1 \lsun\ compared to the observations.  In this model we assumed a mass accretion rate from the disk to the protostar of $\dot{M}_{DtoP} = 0$ \msun\ yr$^{-1}$ in between accretion bursts.  In reality, however, accretion from the disk onto the protostar likely continues at a low rate during the quiescent phases; in their simulations Vorobyov \& Basu (2005b, 2006) found that $\dot{M}_{DtoP}$ is typically $\sim 10^{-8} - 10^{-7}$ \msun\ yr$^{-1}$ during these phases.  Such non-zero values of  $\dot{M}_{DtoP}$, even though they are low, could easily add a few $\times 0.1$ \lsun\ to the total luminosity since this term dominates the total luminosity when present (see \S \ref{discuss_assumptions}).  Thus, we do not consider the excess of time spent at \lbol\ $\sim$ 0.1 \lsun\ in the model to be significant, and we conclude that model 5 best reproduces the observed luminosity distribution of embedded protostars and is the only model that essentially resolves the luminosity problem.

Finally, we note that a K-S test gives a 29\% probability that the observed and model \tbol\ histograms represent the same underlying distribution, comparable to models $1-4$.  As with model 4, model 5 includes a significant population of SEDs with higher \tbol\ than found for embedded sources (35.6\% of the time is spent at \tbol\ $\geq$ 1000 K whereas only 4.5\% of the embedded sources are found at such high \tbol).  Again, this is a consequence of the outflow cavities allowing many lines-of-sight to view direct protostar+disk emission, increasing the $1-100$ \um\ emission and thus the calculated \tbol.  We will return to this point in Section \ref{discuss_evolutionary}.

\section{Discussion}\label{discussion}

\subsection{Scattering and 2-D Geometry}\label{discuss_1to4}

Including the opacity from scattering has an important effect on the results.  Since the opacity from scattering dominates over that from absorption at approximately the same wavelengths over which the protostar+disk input SED peaks ($\sim 0.1-10$ \um), including this opacity significantly increases the total optical depth through the model, causing a corresponding decrease in the amount of emission at these wavelengths escaping from the model.  This reduction in the short-wavelength emission causes a decrease in measured values of \tbol\ (by factors of $\sim 1.5-6$) and slows down the increase in \tbol\ as the model evolves.  As a consequence, the model crosses the Class 0/I boundary in \tbol\ at later times than found by YE05.  We disagree with their conclusion that \tbol\ is not a good evolutionary indicator; we find that it is not quite as good as \lbolsmm\ but remains a satisfactory indicator of evolutionary status (although this will change in the later models; see below).  Despite the important effects of including the opacity from scattering, the general luminosity problem described in \S \ref{intro} remains.

Switching to a 2-D model setup and including a circumstellar disk and rotationally flattened envelope density profile in the model both introduce an inclination dependence, as expected.  However, the dependence is only significant at late times in the collapse of the cores (both because the disk only becomes relatively large and massive at late times, and because the degree of flattening of the envelope starts out very small and increases with time).  Even at such late times the inclination dependence, while introducing spread in the evolution of \lbol, \tbol, and \lbolsmm\ with time, is not large enough to significantly change any of the conclusions (see \S \ref{mod2} and \ref{mod3} for a quantitative description of the changes).  Thus, while important and more physically realistic, these geometry effects do not resolve the discrepancy between the model and the observations.

\subsection{Mass-Loss, Outflow Cavities, and Episodic Accretion:  Connection Between Observational Class and Physical Stage}\label{discuss_evolutionary}

Including the effects of mass-loss, outflow cavities, and episodic accretion has significant effects on the observational signatures of the model.  As shown in both \S \ref{mod4} and \S \ref{mod5}, the variation in \tbol\ and \lbolsmm\ with inclination is dramatically increased by the outflow cavities as lines-of-sight with inclinations less than $\theta_c$ (the cavity semi-opening angle) view direct protostar+disk emission through the cavity in addition to the long-wavelength emission from the envelope.  \tbol\ and \lbolsmm\ calculated from SEDs at the same time in the collapse of the same initial mass core can vary by an order of magnitude or more depending on inclination, and both quantities cross their Class 0/I boundaries at times ranging from very early in the collapse for nearly pole-on lines of sight to very late in the collapse (or even sometimes never for \lbolsmm) for nearly edge-on lines of sight.  This led us to conclude in both \S \ref{mod4} and \S \ref{mod5} that observational Class determined by \tbol\ or \lbolsmm\ is not a good indicator of physical Stage.  To quantify this, Table \ref{tab_fractionstageclass} lists the fraction of total time that model 5 spends in the various Class/Stage combinations according to both \tbol\ and \lbolsmm.  From this table it is apparent that Class determined by \tbol\ only agrees with the underlying physical Stage 40\% of the time; the remaining 60\% is occupied by either \tbol\ measuring Class II while still in the embedded phase, or \tbol\ measuring Class 0 or I while the true physical Stage is the opposite.  \lbolsmm, on the other hand, gives a Class that agrees with the underlying physical Stage 79\% of the time.  Class determined by \lbolsmm\ is thus more likely to trace the true physical stage of an embedded object than if it is determined by \tbol, although we caution that the agreement would be worse if \lbolsmm\ included a defined Class I/II boundary.  We also caution that \lbolsmm\ is more difficult to determine accurately since the calculated value is significantly more sensitive than \tbol\ to the exact wavelengths at which one samples the SED with observations (Dunham et al.~2008).  Thus, in  general, neither evolutionary indicator provides a reliable measure of physical Stage, although \lbolsmm\ is statistically more likely to provide an accurate measure of physical Stage than \tbol.  Furthermore, the entire concept of \tbol\ or \lbolsmm\ increasing monotonically and progressing through the various Classes as the object evolves from Stage 0 through Stage I to Stage II at the end of the embedded phase breaks down once episodic accretion is included.

This point was recognized by Myers et al.~(1998), who acknowledged that the effects of geometry can limit the utility of \tbol.  However, the variation in \tbol\ and \lbolsmm\ with inclination is generally larger in this model than found by previous authors, mainly because we allow $\theta_c$ to increase to very large values (eventually reaching 90\degree\ and ending collapse).  For example, Crapsi et al.~(2008) held $\theta_c$ fixed at 15\degree\ but only presented results for $i > 25$\degree, effectively eliminating any lines-of-sight looking directly down the outflow cavity.  Myers et al.~(1998) concluded that outflow cavities will only change \tbol\ by about a factor of 2 for $\theta_c = 0-25$\degree, which they assumed to be representative of typical cavities, although they did acknowledge the variation will be larger for larger cavities.  Whitney et al.~(2003a, 2003b) assumed small outflow cavities ($\theta_c \sim 20-25$\degree) while Robitaille et al.~(2006) allowed cavities as large as 60\degree; both showed that classification according to infrared colors can vary with inclination but neither discussed the effects on \tbol\ or \lbolsmm.

Are the outflow cavities featured here, with $\theta_c$ increasing to such large values, consistent with observations?  Based on radiative transfer models, Tobin et al.~(2007) found $\theta_c = 15-20$\degree\ for 3 Class 0 sources, and Furlan et al.~(2008) found $\theta_c = 0.1-27$\degree\ for 22 Class I sources.  However, all 25 sources in the combined sample have $\theta_c < i$, and as Furlan et al.~point out, the lack of any sources with large $\theta_c$ could be a selection bias introduced by the fact that such sources may be classified as Class II and thus missing from the sample.  Indeed, some observational evidence for larger outflow cavities is found by Huard et al.~(2006), who measured $\theta_c \sim 50$\degree\ for the Class 0 very low luminosity object (VeLLO; Di Francesco et al.~2007) L1014 from deep near-infrared imaging, and by Arce \& Sargent (2006), who measured $\theta_c \sim 10-60$\degree\ for a sample of 6 Class 0/I embedded sources from millimeter interferometer spectroscopy.

Thus, there is some evidence that outflow cavities can be larger than assumed by Crapsi et al.~(2008) and Myers et al.~(1998).  By choosing an entrainment efficiency between the ejected jet/wind and ambient medium of 100\%, we maximized the speed with which $\theta_c$ increases in order to maximize the effects of mass-loss and outflow cavities and test whether or not episodic accretion is truly needed to resolve the luminosity problem (see below).  However, as discussed in \S \ref{mod4}, even a much smaller, more typically assumed value of 10\% for the entrainment efficiency opens cavities that reach $\theta_c =$ 60, 65, and 75\degree\ at the end of the collapse of the 0.3, 1, and 3 \msun\ initial mass cores, respectively.  In addition to the fact that such a model would lessen the degree to which the luminosity problem is improved (discussed in \S \ref{mod4} and more detail below), it also gives star formation efficiences higher than those found when the assumed entrainment efficiency is 100\%, which we showed in \S \ref{mod4} are generally consistent with estimated values from observations.  More fundamentally, there is a general inconsistency between estimated values of the star formation efficiency ($\sim 30$\%; e.g., Alves et al.~2007; Enoch et al.~2008) and small outflow cavities.  Assuming the cavities reached their maximum sizes immediately after collapse begins, only $13-50$\% of the mass would be removed for $\theta_c = 30-60$\degree.  Simple conservation of momentum and the velocity difference between the jet/wind and outflow argues that the bulk of the mass removed from the system is in the outflow rather than the jet/wind, thus such cavities don't remove enough material to match current estimates of the star formation efficiency.  The assumption of the formation of more than one star per core could help to alleviate this discrepancy, but, on the other hand, outflow cavities don't reach their maximum sizes immediately, making the $13-50$\% an upper limit only.

The outflow cavities create a significant population of embedded objects at high \tbol\ ($\ga 1000$ K).  Is it possible that a population of embedded objects with such high \tbol\ exists?  No such population is included in the Evans et al.~(2009) sample of embedded objects.  This sample is based primarily on the work of Enoch et al.~(2009) and Dunham et al.~(2008); the former compiled a complete sample of embedded objects in Perseus, Serpens, and Ophiuchus while the latter presented the results of a search for all low-luminosity ($\la 1$ \lsun) embedded objects in the molecular clouds and isolated cores observed by the \emph{Spitzer Space Telescope} Legacy Project ``From Molecular Cores to Planet Forming Disks'' (Evans et al.~2003).  While both studies devoted considerable attention to completeness, many of the criteria for identifying candidate embedded objects were based on detections at $24-70$ \um\ and source SEDs exhibiting rising fluxes from shorter to longer wavelengths.  As all panels at $t \geq 34000$ yr in Figure \ref{fig_mod4_seds} clearly show, lines-of-sight passing through outflow cavities and seeing direct protostar+disk emission often do not exhibit rising fluxes at $2-70$ \um\ and do not generally show SEDs typically associated with embedded objects.  It is possible such a population of embedded objects does exist but is not included in the Evans et al.~(2009) Class I sample because of selection biases.  It is difficult to evaluate the exact criteria assumed by Enoch et al.~and Dunham et al.~to determine if the fraction of time the model spends at high \tbol\ would be recovered in their samples since both studies make use of an automatic source classification scheme only available for sources observed by the \emph{Spitzer Space Telescope} (see Evans et al.~2007 for details) and both included significant human judgment to determine what was and was not an embedded source.

Uncertain extinction corrections may also play a role in this discrepancy between observations and models.  When comparing observations of embedded sources to models, corrections must be applied to remove foreground extinction arising from the molecular cloud and ISM (separate from local extinction by the envelope, which will be reradiated in the far-infrared).  Evans et al.~(2009) did correct the 112 embedded sources that are plotted on the BLT diagrams and used to make the histograms in this paper for foreground extinction.  However, in practice it is difficult to determine the value of the foreground extinction to an embedded protostar, so Evans et al.~simply applied the mean extinction to all the Class II objects in the same cloud.  If protostars form in denser parts of clouds, it is possible most embedded sources are undercorrected for foreground extinction.  To quantify this, we took the 1 \msun\ initial mass core at 150,000, extincted the model SEDs by $A_V=10$ and $20$, and then un-extincted them (corrected for extinction) assuming $A_V=5.9$ (the mean value for Perseus, which contributes approximately half of the total sample; see Evans et al.~[2009]).  This decreases \tbol\ for lines-of-sight looking through the outflow cavity from $\sim$3000 K to $\sim$1600 K for $A_V=10$ and $\sim$450 K for $A_V=20$.  Future work must revisit the observational samples and carefully evaluate whether or not a population of embedded sources viewed through outflow cavities and thus exhibiting high \tbol\ and \lbolsmm\ exists, and also whether or not calculated \tbol\ and \lbolsmm\ suffer from an undercorrection for foreground extinction.

\subsection{Mass-Loss, Outflow Cavities, and Episodic Accretion:  Towards Resolving the Luminosity Problem}\label{discuss_luminosity}

The primary motivation for including the modifications to the YE05 model in the step-by-step fashion described in \S \ref{modifications} was to test the hypothesis that episodic accretion is necessary to resolve the luminosity problem and explain the distribution of sources in \lbol\ $-$ \tbol\ space by eliminating other possibilities.  While each had important effects and improved the physical realism of the model, including the scattering from opacity and 2-D effects of a circumstellar disk and rotationally flattened envelope left the conclusions essentially unchanged: the model spent too much time at high \lbol\ ($\ga 1-2$ \lsun) and low \tbol\ ($\la 100-200$ K) compared to that expected from the distribution of embedded sources in \lbol\ $-$ \tbol\ space.

Including the effects of outflow cavities and mass-loss lessened the severity of the luminosity problem but did not eliminate it, even when the effects were maximized by assuming a 100\% entrainment efficiency between the jet/wind ejected by the protostellar system and the surrounding envelope.  A smaller, more typically assumed value of 10\% lessened the degree to which the luminosity problem was resolved (and also increased the star formation efficiencies to values higher than expected from observationally determined estimates).  Thus, even with mass-loss maximized, which minimizes both the protostellar mass and the mass accretion rate and thus minimizes the model luminosity, the model still overpredicts the amount of time spent at \lbol\ $\ga 1-2$ \lsun.  We consider this to be a strong indication of the necessity of invoking episodic accretion to bring models in agreement with observations, on top of the observational and theoretical evidence for such a process described in \S \ref{intro}.

%Indeed, model 5, which includes a simple treatment of episodic mass accretion based on the simulations by Vorobyov \& Basu (2005b, 2006) on top of the other modifications, is the only model that resolves the luminosity problem.  The model does not overpredict the fraction of time spent at high \lbol; rather, it underpredicts it.  This high fraction of time spent at \lbol\ $\sim 0.1$ \lsun\ relative to observations of embedded sources gives a quantitatively worse fit to the observed luminosity distribution, but, as we discuss in \S \ref{mod5}, this ``new luminosity problem'' is likely easily explained by our simple assumption of no mass accretion from the disk onto the protostar during quiescent phases between accretion bursts.  Even a very small accretion rate during such phases, such as found by Vorobyov \& Basu (2005b, 2006), could add a few $\times$ 0.1 \lsun\ to the model and bring the observed and model distributions into better agreement.  Future work must concentrate on incorporating a more detailed treatment of episodic accretion that follows the exact evolution of the Vorobyov \& Basu (2005b, 2006) simulations to fully test this conclusion.

Indeed, model 5, which includes a simple treatment of episodic mass accretion based loosely on the simulations by Vorobyov \& Basu (2005b, 2006) on top of the other modifications, is the only model that essentially resolves the luminosity problem.  If episodic accretion does in fact occur, as supported by the models presented in this paper, there may be important consequences for planet formation since the properties of the circumstellar disk at the end of the embedded stage, in particular the disk mass, will depend on where in the cycle of episodic accretion the system is when the envelope fully dissipates.  Another consequence is that the accretion bursts account for a large fraction of the total accretion onto the protostar:  50\%, 83\%, and 91\% of the final stellar mass accretes during these bursts for the model 5 0.3, 1, and 3 \msun\ initial mass cores.  The range is due almost entirely to different fractions of total time occupied by the FHSC phase (the first 20,000 yr) where mass accretes directly from the envelope onto the protostar.  Thus, if this simple model reflects reality as comparison to observations suggests, between 50-90\% of the final protostellar mass accretes in $\leq 2$\% of the total duration of the embedded phase.  This is in general agreement with Evans et al.~(2009), who used their luminosity distribution of embedded sources and a simple toy model to conclude that 50\% of the final protostellar mass accretes in 7\% of the lifetime.  As they noted, 7\% is an upper limit only since their sample may lack the rarest, most luminous accretion events, a suggestion reinforced by our results.

Finally, we caution that these results do not prove episodic accretion occurs, either as described by Vorobyov \& Basu (2005b, 2006) or in some other fashion.  While we consider the results of this paper to be strong evidence in favor of a process of episodic accretion existing in the formation of low-mass protostars, future work must continue to search for definitive observational evidence that such a process occurs (see discussion in Section 6.4 of Dunham et al.~[2008]).

\subsection{Model Assumptions}\label{discuss_assumptions}

The models presented in this paper are simple, idealized models of star-forming cores that are highly parameterized.  We justified the choice of specific parameterizations and parameter values with theoretical and/or observational constraints in most cases.  The one notable exception is our choice to maximize the momentum entrainment efficiency between the jet/wind and ambient medium in \S \ref{mod4}, and in this case we discussed the effects of varying this parameter.

In general, we consider our results to be robust to different values and parameterizations.  For example, it is the simple presence of a disk in the radiative transfer, rather than the exact disk density profile assumed, that affects our results since any disk-shaped object will introduce a similar inclination dependence in calculated observational signatures.  It is not the exact shape of the outflow cavity (assumed here to follow the streamlines of the collapse solution and thus be conical at large radii) that matters as much as their simple presence in the envelope, as outflow cavities of any shape will increase \lbol, \tbol, and \lbolsmm\ for inclinations that view direct protostar+disk emission through the outflow cavity in addition to the far-infrared and millimeter wavelength emission from the envelope.  It is not the exact choice of burst and quiescent accretion rates or the exact condition upon which a burst begins that matters as much as it is the simple existence of bursts, since a cycle of episodic accretion will, in general, shift the models to lower luminosities except during bursts and will cause periodic increases and decreases in evolutionary indicators like \tbol\ and \lbolsmm.  The details of shape of the model SEDS and the comparison to observations will change, but the overall results will not.

\begin{figure}[hbt!]
\epsscale{0.90}
\plotone{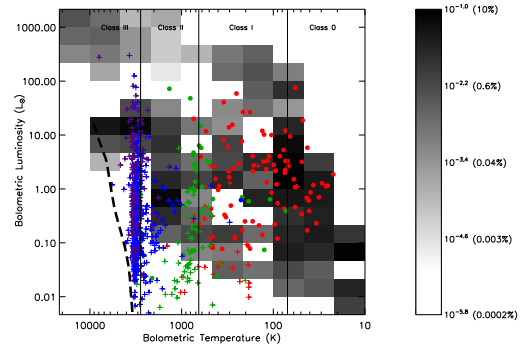}
\caption{\label{fig_mod5a_blt}Same as Figure \ref{fig_mod5a_blt}, except now for the alternate model 5 with $\dot{M}_{DtoP} = 10^{-7}$ \msun\ yr$^{-1}$ in the quiescent accretion phase.  The grayscale pixels indicate the fraction of total time the model spends in each \lbol\ $-$ \tbol\ bin, calculated from Equation \ref{eq_bins}.  The grayscale is displayed in a logarithmic stretch with the scaling chosen to emphasize the full extent of the models in \lbol\ $-$ \tbol\ space.  The mapping between grayscale and fraction of total time is indicated in the legend.  The class boundaries in \tbol\ are taken from Chen et al.~(1995).  The thick dashed line shows the ZAMS (D'Antona \& Mazzitelli 1994) from 0.1 to 2.0 \msun.  The colored symbols show the Young Stellar Objects from Evans et al.~(2009) in this diagram; the colors and symbols hold the same meaning as in Figure \ref{fig_mod1_blt}.}
\end{figure}

To give a quantitative example, motivated by the excess of time spent at low-luminosity ($\sim 0.1$ \lsun) by model 5 compared to the observations as seen in Figure \ref{fig_mod5_hist} and discussed in \S \ref{mod5}, we constructed an alternate version of model 5.  Everything remains the same in this alternate model except $\dot{M}_{DtoP}$, the accretion rate from the disk onto the protostar, is increased from $\dot{M}_{DtoP} = 0$ \msun\ yr$^{-1}$ in the quiescent phases between accretion bursts to $\dot{M}_{DtoP} = 10^{-7}$ \msun\ yr$^{-1}$, in general agreement with Vorobyov \& Basu (2005b, 2006).  The overall results of this change are minor; the peak in time spent by the model at $\sim 0.1$ \lsun\ decreases by about 25\% since the nonzero quiescent $\dot{M}_{DtoP}$ increases the quiescent phase model luminosities, but no substantial changes are introduced in our conclusions.  Figure \ref{fig_mod5a_blt} shows the BLT diagram for this alternate version of model 5; while the overall distribution of time spent in various bins is similar, we note that the location of excluded (white) zones changes.  These excluded zones should not be considered a real effect; slight changes in model parameters can move these zones around without changing the overall model conclusions, and increased sampling of the core mass function beyond three masses (0.3, 1, and 3 \msun) are likely to fill in at least some of them.

A more likely cause of uncertainty than the model parameterizations is in the weighting of the different initial mass cores by the CMF in order to calculate the fractions of total time spent in different \lbol\ and \tbol\ bins for comparison to observations.  The three initial masses (0.3, 1, and 3 \msun) were chosen both by YE05 and by us because they adequately sample both sides of the peak of the CMF ($\sim 1$ \msun).  However, the exact shape of the CMF and thus the relative numbers of different mass cores remains a significant unknown, especially at the low end where many studies suffer from incompleteness effects.  The individual model results are robust to different parameterizations, but the combined comparison to observations is significnatly more uncertain.  This comparison must be revisited as better studies of the CMF become available with new instruments such as SCUBA-II (Ward-Thompson et al.~2007).

Finally, we note that the collapse of the core to form a protostar follows the inside-out collapse of static, singular isothermal spheres first calculated by Shu (1977) and extended by TSC84 to include rotation.  Other collapse solutions that take into account nonzero initial velocities (e.g., Hunter 1977; Fatuzzo et al.~2004) and magnetic fields (e.g., Li \& Shu 1997) tend to increase the accretion rate and would thus worsen the luminosity problem.  On the other hand, Vorobyov \& Basu (2005a) showed that a finite mass reservoir will create a phase of terminally declining accretion rate, an effect included in their collapse simulations featuring episodic accretion (Vorobyov \& Basu 2005b, 2006).  More detailed future models that follow the exact evolution of the Vorobyov \& Basu simulations rather than the simple, idealized models presented here will be needed to fully evaluate the necessity and ability of episodic accretion to resolve the luminosity problem.

\section{Summary}\label{summary}

We have made five modifications to the YE05 evolutionary model in an effort to bring the model in better agreement with observations:  (1) we modified the dust opacities to include isotropic scattering off dust grains (\S \ref{mod1}), (2) we included a circumstellar disk directly in the radiative transfer (\S \ref{mod2}), (3) we included a rotationally flattened envelope density structure following the TSC84 solution for the collapse of slowly rotating cores (\S \ref{mod3}), (4) we included the effects of mass-loss and outflow cavities (\S \ref{mod4}), and (5) we included a simple treatment of episodic mass accretion based on the simulations by Vorobyov \& Basu (2005b, 2006; \S \ref{mod5}).

We find that the first four models all affect the model predictions but are unable to resolve the long-standing luminosity problem.  Including a cycle of episodic accretion, however, can resolve this problem and bring the model predictions in better agreement with observations.  We find that standard assumptions about the interplay between mass accretion and mass loss in our model give star formation efficiencies consistent with recent observations that compare the core mass function (CMF) to the stellar initial mass function (IMF), and that the combination of episodic accretion and increased inclination dependence introduced by the presence of outflow cavities both work to reduce the connection between physical Stage and observational Class as calculated by common evolutionary indicators.  We have outlined future studies needed on both observational and modeling fronts in order to test the conclusions of this paper that episodic accretion is both necessary and sufficient to resolve the luminosity problem.

The authors would like to thank the referee, Alan Boss, for comments that have improved the quality of this paper.  We also thank E. Vorobyov and P. Myers for reading draft versions of this paper and providing helpful comments.  This work is based partly on observations obtained with the \emph{Spitzer Space Telescope}, operated by the Jet Propulsion Laboratory, California Institute of Technology.  This research has made use of NASA's Astrophysics Data System (ADS) Abstract Service.  Support for this work, part of the \emph{Spitzer} Legacy Science Program, was provided by NASA through contracts 1224608, 1288664, 1288658, and RSA 1377304.  Support was also provided by NASA Origins grant NNX 07-AJ72G and NSF grant AST-0607793.  MMD also acknowledges support from a UT Austin University Continuing Fellowship.

\clearpage
%\begin{landscape}
\begin{deluxetable}{lccccccc}
\tabletypesize{\scriptsize}
\tablewidth{0pt}
\tablecaption{\label{tab_fractionperbin}Fraction of Time in BLT Bins}
\tablehead{
\colhead{Bin} & \colhead{Observations} & \colhead{YE05 Model} & \colhead{Model 1} & \colhead{Model 2} & \colhead{Model 3} & \colhead {Model 4} & \colhead{Model 5}}
\startdata
\lbol\ $<$ 0.1 \lsun, 10 $\leq$ \tbol\ $<$ 100 K                &    0\% &  5.4\% &  6.0\% &  6.0\% &  5.5\% &  9.4\% & 19.6\% \\
\lbol\ $<$ 0.1 \lsun, 100 $\leq$ \tbol\ $<$ 1000 K              &  5.4\% &    0\% &    0\% &    0\% &    0\% &  0.2\% &  5.5\% \\
\lbol\ $<$ 0.1 \lsun, 1000 $\leq$ \tbol\ $<$ 10000 K            &  1.8\% &    0\% &    0\% &    0\% &    0\% &  0.6\% &  0.7\% \\
\lbol\ $<$ 0.1 \lsun, \tbol\ $\geq$ 10000 K                     &    0\% &    0\% &    0\% &    0\% &    0\% &    0\% &    0\% \\
0.1 $\leq$ \lbol\ $<$ 10 \lsun, 10 $\leq$ \tbol\ $<$ 100 K      & 32.1\% & 11.5\% & 12.4\% & 20.3\% & 21.3\% & 34.4\% & 28.4\% \\
0.1 $\leq$ \lbol\ $<$ 10 \lsun, 100 $\leq$ \tbol\ $<$ 1000 K    & 43.8\% &  5.8\% &  0.9\% &  0.6\% &  0.6\% & 11.8\% &  8.5\% \\
0.1 $\leq$ \lbol\ $<$ 10 \lsun, 1000 $\leq$ \tbol\ $<$ 10000 K  &  0.9\% &  0.1\% &  0.1\% &  0.8\% &    0\% & 18.3\% & 27.4\% \\
0.1 $\leq$ \lbol\ $<$ 10 \lsun, \tbol\ $\geq$ 10000 K           &    0\% &    0\% &    0\% &    0\% &    0\% &    0\% &    0\% \\
10 $\leq$ \lbol\ $<$ 1000 \lsun, 10 $\leq$ \tbol\ $<$ 100 K     &  2.7\% &  1.0\% & 25.5\% & 26.2\% & 18.7\% &  2.8\% &  1.8\% \\
10 $\leq$ \lbol\ $<$ 1000 \lsun, 100 $\leq$ \tbol\ $<$ 1000 K   & 11.6\% & 58.3\% & 51.2\% & 41.0\% & 53.9\% &  1.2\% &  0.6\% \\
10 $\leq$ \lbol\ $<$ 1000 \lsun, 1000 $\leq$ \tbol\ $<$ 10000 K &  1.8\% & 17.9\% &  3.9\% &  5.1\% &    0\% & 21.2\% &  7.1\% \\
10 $\leq$ \lbol\ $<$ 1000 \lsun, \tbol\ $\geq$ 10000 K          &    0\% &    0\% &    0\% &    0\% &    0\% &    0\% &  0.1\% \\
\lbol\ $\geq$ 1000 \lsun, 10 $\leq$ \tbol\ $<$ 100 K            &    0\% &    0\% &    0\% &    0\% &    0\% &    0\% &    0\% \\
\lbol\ $\geq$ 1000 \lsun, 100 $\leq$ \tbol\ $<$ 1000 K          &    0\% &    0\% &    0\% &    0\% &    0\% &    0\% &    0\% \\
\lbol\ $\geq$ 1000 \lsun, 1000 $\leq$ \tbol\ $<$ 10000 K        &    0\% &    0\% &    0\% &    0\% &    0\% &    0\% &  0.1\% \\
\lbol\ $\geq$ 1000 \lsun, \tbol\ $\geq$ 10000 K                 &    0\% &    0\% &    0\% &    0\% &    0\% &    0\% &  0.2\% \\
\enddata \\
\end{deluxetable}
%\end{landscape}

\begin{deluxetable}{lccccc}
\tabletypesize{\scriptsize}
\tablewidth{0pt}
\tablecaption{\label{tab_models}Summary of Additions to the YE05 Model}
\tablehead{
\colhead{}      & \colhead{}           & \colhead{}         & \colhead{2-D Envelope}    & \colhead{Mass-Loss and}    & \colhead{Episodic} \\
\colhead{Model} & \colhead{Scattering} & \colhead{2-D Disk} & \colhead{Density Profile} & \colhead{Outflow Cavities} & \colhead{Accretion}}
\startdata
Model 1 (\S \ref{mod1}) & Y & N & N & N & N \\
Model 2 (\S \ref{mod2}) & Y & Y & N & N & N \\
Model 3 (\S \ref{mod3}) & Y & Y & Y & N & N \\
Model 4 (\S \ref{mod4}) & Y & Y & Y & Y & N \\
Model 5 (\S \ref{mod5}) & Y & Y & Y & Y & Y \\
\enddata \\
\end{deluxetable}

\begin{deluxetable}{llcc}
\tabletypesize{\scriptsize}
\tablewidth{0pt}
\tablecaption{\label{tab_fractionstageclass}Fraction of Time in Class/Stage Combinations}
\tablehead{
\colhead{} & \colhead{} & \colhead{Stage 0}                    & \colhead{Stage I}                  \\
\colhead{} & \colhead{} & \colhead{$M_{int}/M_{tot} \leq 0.5$} & \colhead{$M_{int}/M_{tot} > 0.5$}}
\startdata
\tbol\    & Class 0 (\tbol\ $< 70$ K)               & 31\% & 6\% \\
          & Class I ($70 \leq$ \tbol\ $\leq 650$ K) & 12\% & 9\% \\
          & Class II (\tbol\ $> 650$ K)             & 10\% & 31\% \\
\hline
\lbolsmm\ & Class 0 (\lbolsmm\ $\leq 175$ K)        & 43\% & 10\% \\
          & Class I (\lbolsmm\ $> 175$ K)           & 11\% & 36\% \\
\enddata \\
\end{deluxetable}

\end{document}